\documentclass[aps,prb,twocolumn,groupedaddress]{revtex4-1}
\usepackage[dvips]{graphicx}
\usepackage{braket}
\usepackage{xcolor}
\begin{document}

\renewcommand{\vec}[1]{\mathbf{#1}}



\title{\textit{Ab initio} studies on the lattice thermal conductivity of silicon clathrate frameworks II and VIII}


\author{Ville J. H\"{a}rk\"{o}nen}
\email[]{ville.j.harkonen@jyu.fi}
\affiliation{Department of Chemistry, University of Jyv\"{a}skyl\"{a}, PO Box 35, FI-40014, Finland}
\author{Antti J. Karttunen}
\email[]{antti.j.karttunen@iki.fi}
\affiliation{Department of Chemistry, Aalto University, FI-00076 Espoo, Finland.}


\date{\today}

\begin{abstract}
The lattice thermal conductivities of silicon clathrate frameworks II and VIII are investigated by using \textit{ab initio} lattice dynamics and iterative solution of the linearized Boltzmann transport equation (BTE) for phonons. Within the temperature range 100-350 K, the clathrate structures II and VIII were found to have lower lattice thermal conductivity values than silicon diamond structure (\textit{d}-Si) by factors of 1/2 and 1/3, respectively. The main reason for the lower lattice thermal conductivity of the clathrate structure II in comparison to \textit{d}-Si was found to be the harmonic phonon spectra, while in the case of the clathrate structure VIII, the difference is mainly due to the harmonic phonon spectra and partly due to shorter relaxation times of phonons. In the studied clathrate frameworks, the anharmonic effects have larger impact on the lattice thermal conductivity than the size of the unit cell. For the structure II, the predicted lattice thermal conductivity differs approximately by factor of 20 from the previous experimental results obtained for a polycrystalline sample at room temperature.
\end{abstract}

\pacs{63.20.D-,63.20.dk, 66.70.Df}

\maketitle

\section{Introduction}
\label{cha:Introduction}

Thermoelectric materials may be used to convert waste heat to electricity and may also be used to improve the energy efficiency of various electronic devices. It is usually considered that crystalline materials with rather high thermoelectric efficiency have rather low lattice thermal conductivities. \cite{Ioffe-SemicondThermoelements-1957,Slack-crtThermoelectrics-1995,Takabatake-RevModPhys.86.669-2014} This has also motivated much of the present research efforts on lattice thermal conductivity, since in order calculate estimates for the thermoelectric efficiency it is usually necessary to calculate the values of lattice thermal conductivity and these numerical results may yield useful information on the mechanisms of lattice thermal conductivity. The calculation of lattice thermal conductivity can be rather challenging task when \textit{ab initio} methods are used, due to the high computational cost of calculating third order (or higher) interatomic forces, for example. 

There are a variety of computational approaches for predicting lattice thermal conductivity, for instance, single mode relaxation time approximations (SMRTs) with Boltzmann transport equation (BTE) \cite{Callaway-PhysRev.113.1046-LatCond-1959,Allen-PhysRevB.88.144302-Improved-Callaway-2013} and iterative solution of linearized BTE. \cite{Omini-iterative-BTE-1995,Omini-PhysRevB.53.9064-iterative-BTE-1996,Broido-PhysRevB.72.014308-LatCond-BTE-2005,Fugallo-PhysRevB.88.045430-Thermal-Cond-2013,Li-shengbte-2014,Chernatynskiy-PhonTS-2015} In the iterative solution of BTE, one obtains, up to some approximation, the non-equilibrium distribution functions dependent on each other, while in the SMRTs the distribution functions are independent (Sec. \ref{Thermalconductivity}). 

One class of crystalline structures for which relatively high thermoelectric efficiencies and low lattice thermal conductivities have been measured is the so-called semiconducting clathrates. \cite{Kasper-24121965.science.clath.1965,Sootsman-thermoelect.-review.2009-ANIE200900598,Christensen.et.al.-2010-B916400F,Takabatake-RevModPhys.86.669-2014} Different mechanisms behind the decreased lattice thermal conductivity in clathrate structures have been proposed. In the compound clathrates such as K$_8$Al$_8$Si$_{38}$, in which the framework is completely or partially filled with guest atoms (K) and the silicon framework includes heteroatoms (Al), the reduction in lattice thermal conductivity is considered to be caused by the increase of phonon-phonon scattering due to the guest atoms. \cite{Nolas-Ge-clath-thermoelect-1998,Cohn-GlasslikeHeatCond-PhysRevLett.82.779-1999,Tadano-ImpactOfRattlers-PhysRevLett.114.095501-2015} For elemental clathrate frameworks with no guest atoms or framework heteroatoms, no such single mechanism has been suggested. For the clathrate framework II (sometimes denoted as Si$_{34}$ or Si$_{136}$), experimental lattice thermal conductivity data have been obtained with polycrystalline samples. For example, the room temperature lattice thermal conductivity of the guest-free clathrate framework II was measured to be 2.5 W/(m K). \cite{Nolas-ThermalCondSi136-2003}

Here we study the lattice thermal conductivity of the silicon diamond (\textit{d}-Si) and the silicon clathrate frameworks II and VIII by combining \textit{ab initio} lattice dynamics with iterative solution of linearized BTE. The solution to the linearized BTE, with 3-phonon and isotopic scattering included, is obtained with the ShengBTE program.\cite{Li-shengbte-2014} The effect of different quantities on the lattice thermal conductivities and the validity of the linearized BTE approach is investigated. The difference in the unit cell size of clathrate structures II and VIII also enables the examination of the effect of unit cell size on the lattice thermal conductivity.

\section{Theory, computational methods and studied structures}
\label{cha:TheoryComputationalMethodsAndStudiedStructures}

\subsection{Lattice dynamics}
\label{LatticeDynamics}
The theory of lattice dynamics discussed in this section has been described, for example, in Refs. \cite{Born-Huang-DynamicalTheory-1954,Maradudin-harm-appr-1971,Maradudin-dyn-prop-solids-1974}. Here, the outline of the procedure which is used to derive the relations implemented in the actual numerical calculations is described. To solve the vibrational states of the crystal, one needs only the $0$th-order wave function \cite{Born-Huang-DynamicalTheory-1954,Maradudin-dyn-prop-solids-1974} and in this approach one assumes that the vibrational Hamiltonian is of the form 
\begin{equation} 
H = T_{n} + \Phi, 
\label{eq:LatticeHamiltonian}
\end{equation}
\begin{equation} 
T_{n} =\frac{1}{2} \sum_{l,\kappa,\alpha} \frac{p^{2}_{\alpha}\left(l \kappa\right)}{M_{\kappa}}, 
\label{eq:KineticEnergy}
\end{equation}
\begin{equation} 
\Phi = \Phi\left[\vec{x}\left(l_{1}\kappa_{1}\right)+\vec{u}\left(l_{1}\kappa_{1}\right),\ldots,\vec{x}\left(l_{n}\kappa_{n}\right)+\vec{u}\left(l_{n}\kappa_{n}\right)\right], 
\label{eq:PotentialEnergy_1}
\end{equation}
\begin{eqnarray} 
\Phi &=& \sum_{n=0} \frac{1}{n!} \sum_{l_{1},\kappa_{1},\alpha_{1}} \cdots \sum_{l_{n},\kappa_{n},\alpha_{n}} \Phi_{\alpha_{1} \cdots \alpha_{n}}\left(l_{1} \kappa_{1};\ldots;l_{n} \kappa_{n} \right) \nonumber \\
			&\times& u_{\alpha_{1}}\left(l_{1} \kappa_{1} \right) \ldots u_{\alpha_{n}}\left(l_{n} \kappa_{n} \right), 
\label{eq:PotentialEnergy}
\end{eqnarray}
\begin{eqnarray} 
&&\Phi_{\alpha_{1} \cdots \alpha_{n}}\left(l_{1} \kappa_{1};\ldots;l_{n} \kappa_{n} \right) \nonumber \\
&&\equiv \left.\frac{\partial^{n}{\Phi}}{\partial{x'_{\alpha_{1}}\left(l_{1} \kappa_{1} \right)} \cdots \partial{x'_{\alpha_{n}}\left(l_{n} \kappa_{n} \right)}  } \right|_{\left\{x'\left(l_{i}\kappa_{i}\right) = x\left(l_{i}\kappa_{i}\right)\right\}},
\label{eq:AtomicForceConstants}
\end{eqnarray}
where $T$ is the kinetic energy and $\Phi$ is the potential energy, $u_{\alpha_{i}}\left(l_{i} \kappa_{i} \right)$ is the displacement of the atom $\kappa_{i}$ in the unit cell labeled by $l_{i}$ from the equilibrium position $\vec{x}\left(l_{i}\kappa_{i}\right)$ in the direction $\alpha_{i}$, $p_{\alpha_{i}}\left(l_{i} \kappa_{i} \right)$ the corresponding momentum and $M_{\kappa_{i}}$ is the atomic mass of atom $\kappa_{i}$. The coefficients $\Phi_{\alpha_{1} \cdots \alpha_{n}}\left(l_{1} \kappa_{1};\ldots;l_{n} \kappa_{n} \right)$ are sometimes called the $n$th order atomic force constants or interatomic force constants. The unit cell index $l$ is defined by the lattice translational vector
\begin{equation} 
\vec{x}\left(l\right) \equiv \vec{x}\left(l_{1},l_{2},l_{3}\right)= l_{1}\vec{a}_{1}+l_{2}\vec{a}_{2}+l_{3}\vec{a}_{3},
\label{eq:LatticeTranslationalVector}
\end{equation}
where $l_{i}$ are integers and the vectors $\vec{a}_{i}$ are called the primitive translational vectors of the lattice. The equilibrium positions can be written as
\begin{equation} 
\vec{x}\left(l\kappa\right) = \vec{x}\left(l\right) + \vec{x}\left(\kappa\right),
\label{eq:EquilibriumPositionVector}
\end{equation}
where $\vec{x}\left(\kappa\right)$ is the position vector of atom $\kappa$ within the unit cell. Sometimes Eq. \ref{eq:LatticeHamiltonian} is written as (static lattice contribution with $n=0$ neglected)
\begin{equation} 
H = H_{0} + H_{A},
\label{eq:LatticeHamiltonian_2}
\end{equation}
with
\begin{eqnarray} 
H_{0} & = & \frac{1}{2} \sum_{l,\kappa,\alpha} \frac{p^{2}_{\alpha}\left(l \kappa\right)}{M_{\kappa}} \nonumber \\
      & + & \frac{1}{2 !}\sum_{l,\kappa,\alpha} \sum_{l',\kappa',\beta} \Phi_{\alpha \beta}\left(l \kappa;l' \kappa' \right) u_{\alpha}\left(l \kappa \right) u_{\beta}\left(l' \kappa' \right),
\label{eq:LatticeHamiltonianHarmonic}
\end{eqnarray}
\begin{eqnarray} 
H_{A} & = & \sum_{n \neq 0,2} \frac{1}{n!} \sum_{l_{1},\kappa_{1},\alpha_{1}} \cdots \sum_{l_{n},\kappa_{n},\alpha_{n}} \Phi_{\alpha_{1} \cdots \alpha_{n}}\left(l_{1} \kappa_{1};\ldots;l_{n} \kappa_{n} \right) \nonumber \\
			&\times& u_{\alpha_{1}}\left(l_{1} \kappa_{1} \right) \ldots u_{\alpha_{n}}\left(l_{n} \kappa_{n} \right), 
\label{eq:LatticeHamiltonianAnharmonic}
\end{eqnarray}
where $H_{0}$ and $H_{A}$ are sometimes called the harmonic and anharmonic vibrational Hamiltonian, respectively.

Within harmonic approximation, only $H_{0}$ is taken into account. One may proceed, for example, by writing the displacements and momenta as a Fourier series (lattice periodicity). After the substitution of the expanded displacements to the potential term one may define the elements of the so-called dynamical matrix as
\begin{equation}
D_{\alpha\beta}\left(\kappa\kappa'|\vec{q}\right) \equiv \sum_{l'} \frac{\Phi_{\alpha\beta}\left(l\kappa;l'\kappa'\right)}{\sqrt{M_{\kappa}M_{\kappa'}}} e^{-i\vec{q}\cdot\left[\vec{x}\left(l\right)-\vec{x}\left(l'\right)\right]},
\label{eq:DynamicalMatrixDef}
\end{equation}
where $\vec{q}$ is the wave vector times $2 \pi$. It can be shown that the dynamical matrix is Hermitian and for Hermitian matrix there exists a set of eigenvectors $\left\{ \vec{e}\left(\kappa|\vec{q}j\right) \right\}$ and eigenvalues $\left\{ \omega_{j}\left(\vec{q}\right) \right\}$ such that
\begin{equation}
\omega^{2}_{j}\left(\vec{q}\right)e_{\alpha}\left(\kappa|\vec{q}j\right) = \sum_{\kappa',\beta}D_{\alpha\beta}\left(\kappa\kappa'|\vec{q}\right)e_{\beta}\left(\kappa'|\vec{q}j\right),
\label{eq:EigenvalueEquation}   
\end{equation}
where $j = 1 \ldots 3n$ is the mode index, $n$ being the number of atoms within unit cell. Sometimes the eigenvalues $\left\{ \omega_{j}\left(\vec{q}\right) \right\}$ are called phonon eigenvalues, phonon frequencies, or as a function of wave vector $\vec{q}$, phonon dispersion relations. The components of the eigenvector $\vec{e}\left(\kappa|\vec{q}j\right)$ of a Hermitian matrix can be chosen to satisfy the orthonormality and closure conditions
\begin{equation}
\sum_{\kappa,\alpha}e_{\alpha}\left(\kappa|\vec{q}j'\right)e^{*}_{\alpha}\left(\kappa|\vec{q}j\right)=\delta_{jj'},
\label{eq:Orthonormality}
\end{equation}
\begin{equation}
\sum_{j}e_{\alpha}\left(\kappa|\vec{q}j\right)e^{*}_{\beta}\left(\kappa'|\vec{q}j\right)=\delta_{\alpha\beta}\delta_{\kappa\kappa'},
\label{eq:Closure}
\end{equation}
where $\delta_{\alpha\beta}$ is the Kronecker delta. In addition to the Fourier series expansion, one may introduce another unitary transform for the Fourier coefficients in terms of the so-called normal coordinates and eigenvectors $\vec{e}\left(\kappa|\vec{q}j\right)$. Furthermore, one may write the normal coordinates in terms of the so-called creation $\hat{a}^{\dagger}_{\vec{q}j}$ and annihilation operators $\hat{a}_{\vec{q}j}$. The mentioned transforms are included in the following expansions of displacement and momentum (which are considered now as operators) 
\begin{eqnarray} 
\hat{u}_{\alpha}\left(l \kappa\right) & = & \left(\frac{\hbar}{2 N M_{\kappa}}\right)^{1/2} \sum_{\vec{q},j} \omega^{-1/2}_{j}\left(\vec{q} \right) e^{i \vec{q}\cdot \vec{x}\left(l\right)} e_{\alpha}\left(\kappa| \vec{q} j\right) \nonumber \\
                                      & \times & \left( \hat{a}_{\vec{q}j} + \hat{a}^{\dagger}_{-\vec{q}j} \right),
\label{eq:DisplacementExpansion}
\end{eqnarray}
\begin{eqnarray} 
\hat{p}_{\alpha}\left(l \kappa\right) & = & -i \left(\frac{\hbar M_{\kappa}}{2 N }\right)^{1/2} \sum_{\vec{q},j} \omega^{1/2}_{j}\left(\vec{q} \right) e^{i \vec{q}\cdot \vec{x}\left(l\right)} e_{\alpha}\left(\kappa| \vec{q} j\right) \nonumber \\
                                      & \times & \left( \hat{a}_{\vec{q}j} - \hat{a}^{\dagger}_{-\vec{q}j} \right),
\label{eq:MomentumExpansion}
\end{eqnarray}
where $N$ is the number of $\vec{q}$-points in the $\vec{q}$-mesh. After the substitution of expansion in Eqs. \ref{eq:DisplacementExpansion} and \ref{eq:MomentumExpansion} to $H_{0}$ given by Eq. \ref{eq:LatticeHamiltonianHarmonic}, one can write
\begin{equation} 
\hat{H}_{0}=\sum_{\vec{q},j}\hbar \omega_{j} \left(\vec{q}\right) \left(\frac{1}{2}+ \hat{a}^{\dagger}_{\vec{q}j} \hat{a}_{\vec{q}j} \right),
\label{eq:HarmonicCrystalHamiltonianSecondQuant}
\end{equation}
The anharmonic Hamiltonian operator $\hat{H}_{A}$ can be obtained by the substitution of expansion in Eq. \ref{eq:DisplacementExpansion} to Eq. \ref{eq:LatticeHamiltonianAnharmonic} and can be written as (Ref. \cite{Maradudin-dyn-prop-solids-1974}, p. 32)
\begin{eqnarray} 
&&\hat{H}_{A} = \sum_{\vec{q},j} V\left(\vec{q}j\right) \hat{A}_{\vec{q}j} +\sum_{n=3} \sum_{\vec{q}_{1},j_{1}} \cdots \sum_{\vec{q}_{n},j_{n}} \nonumber \\
&&\times V\left(\vec{q}_{1}j_{1};\ldots;\vec{q}_{n}j_{n}\right) \hat{A}_{\vec{q}_{1},j_{1}}	\cdots \hat{A}_{\vec{q}_{n},j_{n}},
\label{eq:LatticeHamiltonianAnharmonicSecondQuantEq_1} 
\end{eqnarray}
where
\begin{equation} 
\hat{A}_{\vec{q}j} =   \hat{a}_{\vec{q}j} + \hat{a}^{\dagger}_{-\vec{q}j},
\label{eq:CondensedNotatCreAnnihil}
\end{equation}
and \cite{Born-Huang-DynamicalTheory-1954}
\begin{eqnarray} 
&&V\left(\vec{q}_{1}j_{1};\ldots;\vec{q}_{n}j_{n}\right) \nonumber \\
&&= \frac{1}{n!} \left(\frac{\hbar}{2 N}\right)^{n/2} \frac{N \Delta\left( \vec{q}_{1} +\cdots+\vec{q}_{n} \right) }{ \left[\omega_{j_{1}}\left(\vec{q}_{1} \right) \cdots \omega_{j_{n}} \left(\vec{q}_{n} \right) \right]^{1/2} } \nonumber \\
&&\times \sum_{\kappa_{1},\alpha_{1}} \sum_{l_{2},\kappa_{2},\alpha_{2}} \cdots \sum_{l_{n},\kappa_{n},\alpha_{n}} \Phi_{\alpha_{1} \cdots \alpha_{n}}\left(0 \kappa_{1};l_{2} \kappa_{2};\ldots;l'_{n} \kappa'_{n} \right) \nonumber \\
&&\times \frac{e_{\alpha_{1}}\left(\kappa_{1}|\vec{q}_{1}j_{1}\right)}{M^{1/2}_{\kappa_{1}}} \cdots \frac{e_{\alpha_{n}}\left(\kappa_{n}|\vec{q}_{n}j_{n}\right)}{M^{1/2}_{\kappa_{n}}} e^{i \left[\vec{q}_{2}\cdot \vec{x}\left(l_{2}\right)+\cdots+\vec{q}_{n}\cdot \vec{x}\left(l_{n}\right) \right]}. \nonumber \\
\label{eq:LatticeHamiltonianAnharmonicSecondQuantEq_2} 
\end{eqnarray}
In the present work, the linear term in Eq. \ref{eq:LatticeHamiltonianAnharmonicSecondQuantEq_1} is neglected, since it vanishes by the assumption that forces on atoms at equilibrium vanish. However, in some applications, such as in the case of homogenously deformed lattice, the linear term cannot be neglected in general. \cite{Leibfried-SolidStatePhysics-1961}

To obtain the eigenvalues $\left\{ \omega_{j}\left(\vec{q}\right) \right\}$ within the harmonic approximation, one needs to know the second order interatomic forces $\left\{ \Phi_{\alpha\beta}\left(l\kappa;l'\kappa'\right) \right\}$ and the eigenvectors $\left\{ \vec{e}\left(\kappa|\vec{q}j\right) \right\}$. If the anharmonicity in the system is rather small ($\hat{H}_{A}$  sufficiently smaller than $\hat{H}_{0}$), then the anharmonic Hamiltonian may be considered as a perturbation on the harmonic one. In actual calculations, one usually imposes periodic boundary conditions and considers a finite mesh of wave vectors when Eq. \ref{eq:EigenvalueEquation} is used to obtain the eigenvalues. The calculation of the atomic force constants for higher orders than $3$ or $4$ is a rather challenging task from the computational point of view.

\subsection{Thermal conductivity}
\label{Thermalconductivity}
The heat flux $J$ and thermal conductivity $\kappa$ are related as 
\begin{equation}
J_{\alpha} = - \sum_{\beta} \kappa_{\alpha \beta} \frac{\partial{T}}{\partial{x_{\beta}}},
\label{eq:HeatFluxConductivity}
\end{equation}
which is a phenomenological relation. In the BTE approach, one usually assumes that the heat flux is of the form 
\begin{equation}
\vec{J} = \frac{1}{V} \sum_{\vec{q},j} \hbar \omega_{j}\left(\vec{q}\right) \vec{v}\left(\vec{q}j\right) n_{\vec{q}j}.
\label{eq:BTEHeatFlux}
\end{equation}
where $\vec{v}\left(\vec{q}j\right)$ is the phonon group velocity for the state labeled by $\vec{q}j$, $V$ is the volume of the unit cell and $n_{\vec{q}j}$ is the non-equilibrium phonon distribution function. The group velocity can be written as
\begin{equation}
v_{\alpha}\left(\vec{q}j\right) = \frac{\partial{ \omega_{j}\left(\vec{q}\right) }}{\partial{q_{\alpha}}}.
\label{eq:GroupVelocity}
\end{equation}
More general forms of the energy flux (including heat flux) have been derived and it has been shown that Eq. \ref{eq:BTEHeatFlux} is a special case of this more general energy flux. \cite{Hardy-PhysRev.132.168-Energy-Flux-1963} The methods used to obtain the numerical results reported in this work are based on Eq. \ref{eq:BTEHeatFlux}. In Eq. \ref{eq:BTEHeatFlux}, all the other quantities may be obtained by using the harmonic approximation, but $n_{\vec{q}j}$ is unknown and the BTE approach can be used to obtain estimates for it. To obtain estimates for $n_{\vec{q}j}$, it is assumed that the deviations from equilibrium distribution $\bar{n}_{\vec{q}j}$ are relatively small and are caused by the temperature gradients. Thus, one may expand $\bar{n}_{\lambda}$ up to first order as \cite{Ziman-ElectronsPhonons-1960,Srivastava-PhysicsOfPhonons-1990,Omini-PhysRevB.53.9064-iterative-BTE-1996} (notation $\vec{q}j \rightarrow \lambda$)  
\begin{eqnarray}
&&n_{\lambda} \approx \bar{n}_{\lambda}\left(\hbar \omega_{\lambda}  + \vec{F}_{\lambda} \cdot \nabla T\right) = \frac{1}{e^{\left(\hbar \omega_{\lambda} + \vec{F}_{\lambda} \cdot \nabla T\right) \beta } -1} \nonumber \\
&&\approx \bar{n}_{\lambda} + \left.\frac{\partial{\bar{n}_{\lambda}}}{\partial{\hbar \omega'_{\lambda}}} \right|_{\omega'_{\lambda} = \omega_{\lambda}} \vec{F}_{\lambda} \cdot \nabla T, 
\label{eq:BoseEinsteinDeviationEq_1}
\end{eqnarray}
which can be written as
\begin{equation}
n_{\lambda} \approx \bar{n}_{\lambda} - \beta \bar{n}_{\lambda} \left(\bar{n}_{\lambda} + 1\right) \vec{F}_{\lambda} \cdot \nabla T. 
\label{eq:BoseEinsteinDeviationEq_2}
\end{equation}
In Eqs. \ref{eq:BoseEinsteinDeviationEq_1} and \ref{eq:BoseEinsteinDeviationEq_2}, $\vec{F}_{\lambda}$ measures the deviation of energy from the equilibrium value with units $energy \times length/ temperature$ and $\beta = 1/k_{B}T$. Substitution of Eq. \ref{eq:BoseEinsteinDeviationEq_2} to Eq. \ref{eq:BTEHeatFlux} gives (first term on the right hand side of Eq. \ref{eq:BoseEinsteinDeviationEq_2} vanishes)
\begin{equation}
J_{\gamma} \approx -\frac{\beta}{V} \sum_{\lambda}  \sum^{3}_{\alpha = 1} \hbar \omega_{\lambda} v_{\gamma}\left(\lambda\right)  \bar{n}_{\lambda} \left(\bar{n}_{\lambda} + 1\right)  F_{\alpha,\lambda} \frac{\partial{T}}{\partial{x_{\alpha}}},
\label{eq:BTEHeatFluxEq_2}
\end{equation}
and a comparison of Eqs. \ref{eq:HeatFluxConductivity} and \ref{eq:BTEHeatFluxEq_2} shows that
\begin{equation}
\kappa_{\alpha \beta} = \frac{ \hbar }{k_{B}T V} \sum_{\lambda}   \omega_{\lambda} v_{\alpha}\left(\lambda\right) \bar{n}_{\lambda} \left(\bar{n}_{\lambda} + 1\right)  F_{\beta,\lambda}.
\label{eq:HeatFluxConductivityEq_2}
\end{equation}
The unknown quantity $\vec{F}_{\lambda}$ is solved iteratively from linearized BTE, for example, by using ShengBTE.

In this approach BTE, for example for phonons, can be written as \cite{Ziman-ElectronsPhonons-1960,Srivastava-PhysicsOfPhonons-1990}
\begin{equation}
\frac{\partial{n_{\lambda}}}{\partial{t}}= \left.\frac{\partial{n_{\lambda}}}{\partial{t}}\right|_{col} + \left.\frac{\partial{n_{\lambda}}}{\partial{t}}\right|_{diff} + \left.\frac{\partial{n_{\lambda}}}{\partial{t}}\right|_{ext}.
\label{eq:BTEGeneral_1}
\end{equation}
where $n_{\lambda}$ is the non-equilibrium distribution function for phonons and the rate of change of $n_{\lambda}$ is written as a sum of collision (col), diffusion (diff) and external field (ext) terms. In the absence of external electric and magnetic fiels and assuming the steady state condition, one may approximate Eq. \ref{eq:BTEGeneral_1} as
\begin{equation}
\left.\frac{\partial{n_{\lambda}}}{\partial{t}}\right|_{col} + \left.\frac{\partial{n_{\lambda}}}{\partial{t}}\right|_{diff} = 0, \quad \frac{\partial{n_{\lambda}}}{\partial{t}} = \left.\frac{\partial{n_{\lambda}}}{\partial{t}}\right|_{ext} = 0.
\label{eq:BTEGeneral_2}
\end{equation}
Furthermore, one approximates \cite{Ziman-ElectronsPhonons-1960,Srivastava-PhysicsOfPhonons-1990}
\begin{equation}
\left.\frac{\partial{n_{\lambda}}}{\partial{t}}\right|_{diff} \approx -\vec{v}\left(\lambda\right) \cdot \nabla T \frac{\partial{\bar{n}_{\lambda}}}{\partial{T}}.
\label{eq:BTEApproxDiffusion_2}
\end{equation}
Thus, it is assumed that a local equilibrium exists at different parts of the crystal at different temperatures.

The collision term can be obtained as follows. In general, the probability for a process in which a phonon labeled by $\lambda$ vanishes and two phonons $\lambda',\lambda''$ are created is proportional to \cite{Dirac-PrinciplesOfQM-1958}
\begin{equation}
m_{\lambda} \left(m_{\lambda'} + 1\right) \left(m_{\lambda''} + 1\right),
\label{eq:BTEPhononProbability_1}
\end{equation}
where $m_{\lambda}$ etc. are number of phonons on the particular state. For an opposite process
\begin{equation}
m_{\lambda} m_{\lambda'} \left(m_{\lambda''} + 1\right).
\label{eq:BTEPhononProbability_2}
\end{equation}
In the expressions of transition probabilities, there usually is some factors ensuring the conservation of energy and possible momentum and these are included in the following expressions
\begin{equation}
m_{\lambda} \left(m_{\lambda'} + 1\right) \left(m_{\lambda''} + 1\right)P^{\lambda' \lambda''}_{\lambda}, 
\label{eq:BTEPhononProbability_3}
\end{equation}
\begin{equation}
m_{\lambda} m_{\lambda'} \left(m_{\lambda''} + 1\right)P^{\lambda''}_{\lambda \lambda'}.
\label{eq:BTEPhononProbability_4}
\end{equation}
Thus, the total rate of change of the distribution function of a state $\lambda$ can be written as
\begin{eqnarray}
&&\left.\frac{\partial{n_{\lambda}}}{\partial{t}}\right|_{col} = \sum_{\lambda',\lambda''} \left[n_{\lambda} n_{\lambda'} \left(n_{\lambda''} + 1\right)P^{\lambda''}_{\lambda \lambda'}  \right.\nonumber \\
&&-\left. n_{\lambda} \left(n_{\lambda'} + 1\right) \left(n_{\lambda''} + 1\right)P^{\lambda' \lambda''}_{\lambda}\right].
\label{eq:BTEPhononProbability_5}
\end{eqnarray}
In the actual calculation of the total rate of change in Eq. \ref{eq:BTEPhononProbability_5}, all the combinations of the matrix elements of the form (all different 3-phonon processes)
\begin{equation}
\left|\braket{n|\hat{A}_{\lambda}\hat{A}_{\lambda'}\hat{A}_{\lambda''}|m}\right|^{2},
\label{eq:BTEPhononProbability_6}
\end{equation}
must be considered. In Eq. \ref{eq:BTEPhononProbability_6}, $\ket{n}$ and $\ket{m}$ are eigenkets of the harmonic Hamiltonian $\hat{H}_{0}$. By writing all the different combinations obtained from Eq. \ref{eq:BTEPhononProbability_6}, writing the total scattering rate in a similar way as in Eq. \ref{eq:BTEPhononProbability_5}, taking a formal non-equilibrium ensemble average, substituting for all distributions the approximation given by Eq. \ref{eq:BoseEinsteinDeviationEq_2}, neglecting the terms $\vec{F}_{\lambda} \cdot \nabla T$ with higher powers than one, and finally substituting the result to BTE given by Eq. \ref{eq:BTEGeneral_2}, one may write for the $\alpha$th component (possible numerical factors and equilibrium distribution functions are absorbed into the expressions of transition probabilities $\Gamma^{\lambda''}_{\lambda \lambda'}, \Gamma^{\lambda' \lambda''}_{\lambda }$)
\begin{eqnarray}
&&\sum_{\lambda'} \sum_{\lambda''} \left( \Gamma^{\lambda''}_{\lambda \lambda'} + \Gamma^{\lambda' \lambda''}_{\lambda }  \right) F_{\alpha,\lambda}  \frac{\partial{T}}{\partial{x_{\alpha}}} \nonumber \\
&&+  \sum_{\lambda'} \sum_{\lambda''} \left[  \Gamma^{\lambda''}_{\lambda \lambda'} \left(-F_{\alpha,\lambda''} +F_{\alpha,\lambda'} \right) \right.\nonumber \\
&&+ \left. \Gamma^{\lambda' \lambda''}_{\lambda}  \left(-F_{\alpha,\lambda'}-F_{\alpha,\lambda''} \right) \right] \frac{\partial{T}}{\partial{x_{\alpha}}} \nonumber \\
&& = v_{\alpha}\left(\lambda\right) \frac{\partial{T}}{\partial{x_{\alpha}}} \left[\frac{ \hbar \omega_{\lambda} }{ T } \bar{n}_{\lambda} \left(\bar{n}_{\lambda} + 1\right)\right]. 
\label{eq:BTETotal_1}
\end{eqnarray}  
Equation \ref{eq:BTETotal_1} can be used to solve the quantity $F_{\alpha,\lambda}$ and it can be written as
\begin{eqnarray}
&&F_{\alpha,\lambda} = \frac{1}{X_{\lambda}} \sum_{\lambda'} \sum_{\lambda''} \left[  \Gamma^{\lambda''}_{\lambda \lambda'} \left(F_{\alpha,\lambda''} -F_{\alpha,\lambda'} \right) \right. \nonumber \\
&&+ \left. \Gamma^{\lambda' \lambda''}_{\lambda }  \left(F_{\alpha,\lambda'}+F_{\alpha,\lambda''} \right) \right] \nonumber \\
&&+ \frac{ \hbar \omega_{\lambda} v_{\alpha}\left(\lambda\right) }{ T X_{\lambda} } \bar{n}_{\lambda} \left(\bar{n}_{\lambda} + 1\right), 
\label{eq:BTETotal_2}
\end{eqnarray}  
with
\begin{equation}
X_{\lambda} \equiv \sum_{\lambda'} \sum_{\lambda''} \left( \Gamma^{\lambda''}_{\lambda \lambda'} + \Gamma^{\lambda' \lambda''}_{\lambda }  \right).
\label{eq:BTETotal_3}
\end{equation}  
The transition probabilities $\Gamma^{\lambda''}_{\lambda \lambda'},\Gamma^{\lambda' \lambda''}_{\lambda }$ can be obtained, for example, from the golden rule applied to the 3rd order anharmonic Hamiltonian \cite{Ward-PhysRevB.80.125203-Tcond-2009} (ensemble averaged with the harmonic Hamiltonian)  or from the phonon self energy. \cite{Maradudin-Fein-ScatteringOfNeutrons-1962,Li-shengbte-2014} For instance, by using the golden rule
\begin{equation}
\Gamma = \frac{2 \pi}{\hbar}  \left|\braket{n|\hat{H}_{3}|m}\right|^{2} \delta\left( E_{n} - E_{m} \right), 
\label{eq:BTEGoldenRule_1}
\end{equation}  
expressions like the following can be obtained (by taking a canonical ensemble average from Eq. \ref{eq:BTEGoldenRule_1})
\begin{eqnarray}
&&\Gamma^{\lambda''}_{\lambda \lambda'} = \frac{2 \pi}{\hbar^{2} } \sum_{\lambda'} \sum_{\lambda''} \left|V\left(\lambda;\lambda';\lambda''\right)\right|^{2} \nonumber \\
&&\times \bar{n}_{\lambda} \bar{n}_{\lambda'}  \left( \bar{n}_{\lambda''} + 1\right) \delta\left( \omega_{\lambda} + \omega_{\lambda'} - \omega_{\lambda''} \right)
\label{eq:BTEGoldenRule_2}
\end{eqnarray}  
If the isotopic scattering is included, one usually adds the contribution of the isotopic scattering rate to $X_{\lambda}$ \cite{Ward-PhysRevB.80.125203-Tcond-2009,Li-shengbte-2014} and thus assumes that different scattering mechanisms are independent. Sometimes one denotes \cite{Ward-PhysRevB.80.125203-Tcond-2009}
\begin{equation}
F^{0}_{\alpha,\lambda} \equiv \frac{ \hbar \omega_{\lambda} v_{\alpha}\left(\lambda\right) }{ T X_{\lambda} } \bar{n}_{\lambda} \left(\bar{n}_{\lambda} + 1\right),
\label{eq:BTETotal_4}
\end{equation}  
which is the solution of Eq. \ref{eq:BTETotal_2} with 
\begin{equation}
F_{\alpha,\lambda'} = F_{\alpha,\lambda''} = 0,
\label{eq:BTETotal_5}
\end{equation}  
or when
\begin{eqnarray}
&&\sum_{\lambda'} \sum_{\lambda''} \left[ \Gamma^{\lambda''}_{\lambda \lambda'} \left(F_{\alpha,\lambda''} -F_{\alpha,\lambda'} \right) \right. \nonumber \\
&&+\left. \Gamma^{\lambda' \lambda''}_{\lambda}  \left(F_{\alpha,\lambda'}+F_{\alpha,\lambda''} \right) \right] = 0.
\label{eq:BTETotal_6}
\end{eqnarray}

By defining the relaxation time (RT) \cite{Ward-PhysRevB.80.125203-Tcond-2009}
\begin{equation}
\tau_{\alpha}\left(\lambda\right) \equiv \frac{T F_{\alpha,\lambda}}{ \hbar \omega_{\lambda} v_{\alpha}\left(\lambda\right)},
\label{eq:GeneralRelaxTimeDef}
\end{equation}
and using the expression for the heat capacity at constant volume
\begin{equation}
c_{v}\left(\lambda\right) = k_{B} \beta^{2} \hbar^{2}\omega^{2}_{\lambda} \bar{n}_{\lambda} \left(\bar{n}_{\lambda} + 1\right),
\label{eq:HeatCapasityConstVolume}
\end{equation}
one may write Eq. \ref{eq:HeatFluxConductivityEq_2} as 
\begin{equation}
\kappa_{\alpha \beta} = \frac{1}{V} \sum_{\lambda}  v_{\alpha}\left(\lambda\right) v_{\beta}\left(\lambda\right) c_{v}\left(\lambda\right) \tau_{\beta}\left(\lambda\right),
\label{eq:DiagonalThermalConductivity}
\end{equation}
and for crystals with cubic symmetry 
\begin{equation}
\kappa_{\alpha \alpha} = \frac{1}{ V} \sum_{\lambda}  v^{2}_{\alpha}\left(\lambda\right) c_{v}\left(\lambda\right) \tau_{\alpha}\left(\lambda\right).
\label{eq:DiagonalThermalConductivityCubic}
\end{equation}
The 0th order relaxation time
\begin{equation}
\tau^{0}_{\alpha}\left(\lambda\right) = \frac{ \bar{n}_{\lambda} \left(\bar{n}_{\lambda} + 1\right) }{  X_{\lambda} } = \frac{T F^{0}_{\alpha,\lambda}}{ \hbar \omega_{\lambda} v_{\alpha}\left(\lambda\right)},
\label{eq:ZerothOrderRelaxTime}
\end{equation}
is sometimes called the single mode relaxation time (SMRT) since the deviations $\vec{F}_{\lambda'},\vec{F}_{\lambda''}$ vanish (Eq. \ref{eq:BTETotal_5}), while $\vec{F}_{\lambda}$ deviates from its equilibrium value by an amount given by Eq. \ref{eq:BTETotal_4}.

The connection of $\vec{F}_{\lambda},n_{\lambda}$ and $\bar{n}_{\lambda}$ can be obtained from Eq. \ref{eq:BoseEinsteinDeviationEq_2}
\begin{equation}
- \vec{F}_{\lambda} \cdot \nabla T = \frac{k_{B} T}{\bar{n}_{\lambda} + 1 } \left( \frac{n_{\lambda}}{\bar{n}_{\lambda}} - 1 \right). 
\label{eq:BoseEinsteinDeviationEq_22}
\end{equation}
It can be seen from Eq. \ref{eq:BTETotal_2} that $\vec{F}_{\lambda}$ is independent of temperature gradient. Since the steady state condition is assumed in the derivation of phonon BTE (local temperatures exist), one may assume that the temperature gradient is constant (independent of time) and has a negative value, that is, temperature decreases in the positive direction. Thus, one may write Eq. \ref{eq:BoseEinsteinDeviationEq_22} as
\begin{equation}
T_{G} \sum_{\alpha} F_{\alpha,\lambda} = \frac{k_{B} T}{\bar{n}_{\lambda} + 1 } \left( \frac{n_{\lambda}}{\bar{n}_{\lambda}} - 1 \right), 
\label{eq:BoseEinsteinDeviationEq_23}
\end{equation}
where $T_{G}$ is now some positive constant. It follows from Eq. \ref{eq:BoseEinsteinDeviationEq_23} that for
\begin{equation}
\sum_{\alpha} F_{\alpha,\lambda} < 0 \Rightarrow n_{\lambda} < \bar{n}_{\lambda},
\label{eq:BoseEinsteinDeviationEq_24}
\end{equation}
and vice versa. This means that $\tau_{\alpha}\left(\lambda\right)$ can in general also have negative values, which might result in negative thermal conductivity values calculated from Eq. \ref{eq:DiagonalThermalConductivity} for some particular states $\lambda$, implying that the heat flux is positive towards higher temperatures for these states.

In the calculation of the lattice thermal conductivity, ShengBTE does not take into account the shifts of phonon eigenvalues due to third and fourth order atomic force constants. In particular, the fourth order atomic force constants may have some significance since quartic contribution to frequency shift can be obtained from first order phonon self-energy \cite{Maradudin-Fein-ScatteringOfNeutrons-1962} (or some other perturbation method), while the lowest order contributions to frequency shift due to third order atomic forces are in the second order. In ShengBTE, the so-called phase space calculations for the allowed processes (delta functions for momentum and energy in Eq. \ref{eq:BTEGoldenRule_2}) are made by using the harmonic phonon eigenvalues $\omega_{\lambda}$, assumed to be independent of temperature. In addition, the third and fourth order forces change the eigenvectors obtained within the harmonic approximation and the thermal expansion shifts the phonon eigenvalues as a function of temperature. \cite{Maradudin-ThermalExpFreqShifts-1962} These effects are neglected in the present approach. Thus, if the interactions in the system are sufficiently strong, the harmonic approximation may not describe the system as expected and more rigorous methods may be needed to describe the system in more proper detail. Techniques based on the variational method have already been used to calculate lifetimes and cross sections for materials which are experimentally known to have rather strong anharmonicity. \cite{Paulatto-PhysRevB.91.054304_lifetimes-2015}

\subsection{Studied structures and computational details}
\label{StudiedStructuresAndComputationalDetails}

\begin{figure}
\includegraphics[width=0.47\textwidth]{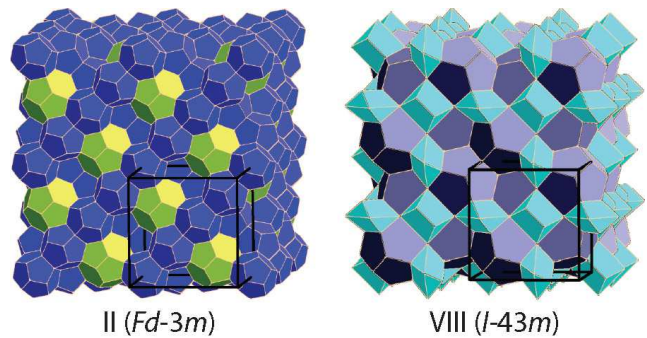}
\caption{Schematic figures of the silicon clathrate frameworks II and VIII studied in this work. The vertices of the polyhedral cages represent silicon atoms. The crystallographic unit cell edges are drawn in black. For a more detailed description of the framework structures, see \cite{Karttunen-Structuralprinc-2010} (Color online).}
\label{fig:II_VIII_structure}
\end{figure}

The structural characteristics of the silicon structures considered in this paper are described in Table \ref{tab:parameters} and the clathrate frameworks II and VIII are illustrated in Fig \ref{fig:II_VIII_structure}. 
\begin{table*}
\caption{Structural data and computational details for the studied structures.}
		\begin{tabular}{cccccccc}
		\hline\hline
		Structure                                  & Space Group                  & Atoms/cell\footnote[1]{number of atoms in primitive cell.} & Elect. (k$_{1}$,k$_{2}$,k$_{3}$)\footnote[2]{The mesh used for the electronic $\vec{k}$-sampling.} &  Phon. (q$_{1}$,q$_{2}$,q$_{3}$)\footnote[3]{$\vec{q}$-meshes for phonon calculations, phonon density-of-states calculations, and lattice thermal conductivity calculations, respectively.} & PDOS (q$_{1}$,q$_{2}$,q$_{3}$)$^c$ & TCOND (q$_{1}$,q$_{2}$,q$_{3}$)$^c$ & $a$(\AA)\footnote[4]{Lattice constant of the optimized primitive unit cell used in the calculations.} \\\hline
	  \textit{d}-Si                              & $Fd\bar{3}m\left(227\right)$ & 2                &  12,12,12                              & 8,8,8                                  & 80,80,80 & 24,24,24 & 5.47 \\ 
		II                                         & $Fd\bar{3}m\left(227\right)$ & 34               &   6,6,6                                & 4,4,4                                  & 30,30,30 & 8,8,8 & 14.74  \\
		VIII                                       & $I\bar{4}3m\left(217\right)$ & 23               &   6,6,6                                & 4,4,4                                  & 40,40,40 & 10,10,10 & 10.10 \\
		\hline\hline
		\end{tabular}
		\label{tab:parameters}
\end{table*}

The \textit{ab initio} density functional calculations to optimize the structures and to calculate the phonon dispersion relations were carried out with the Quantum Espresso program package (QE, version 5.0.3).\cite{QE-2009} The silicon atoms were described using ultrasoft pseudopotentials and plane wave basis set\cite{Garrity-pseudopotentials-2014}. The Generalized Gradient Approximation (GGA) was applied by using the PBE exchange-correlation energy functionals. \cite{Perdew-generalized-1996} In all calculations, the following kinetic energy cutoffs have been used: 44 Ry for wavefunctions and 352 Ry for charge densities and potentials. The applied $\vec{k}$- and $\vec{q}$-sampling for each studied structure are listed in Table \ref{tab:parameters}. The $\vec{q}$-meshes for the lattice thermal conductivity calculations \textit{et cetera} were Fourier interpolated from the mesh used in the corresponding phonon calculation (QE module matdyn.x). \cite{Baroni-RevModPhys.73.515-DFTP-2001} We carried out convergence tests for both total energy and phonon calculations with different $\vec{k}$-meshes for \textit{d}-Si. The convergence with (12,12,12) $\vec{k}$-mesh was accepted and thus used. Similarly, an acceptable convergence was found for the (8,8,8) $\vec{q}$-mesh in the case of \textit{d}-Si. The $\vec{k}$- and $\vec{q}$-meshes listed in in Table \ref{tab:parameters} for the clathrate frameworks were chosen as a compromise between accuracy and computational cost. Both the lattice constants and the atomic positions of the studied structures were fully optimized (applying the space group symmetries listed in Table \ref{tab:parameters}, in which the lattice constants for the optimized structures are also shown). In the structural optimizations, the convergence thresholds on total energy and forces were set to 10$^{-6}$ a.u. and 10$^{-5}$ a.u., respectively. The non-analytic corrections to dynamical matrices in the limit $\vec{q} \rightarrow 0$ were taken into account in the QE and ShengBTE calculations. All lattice thermal conductivity calculations were made by using the version v1.0.2 of ShengBTE.

To our knowledge, there is no way to label the phonon modes (and phonon eigenvectors) uniquely at a point of degeneracy when the diagonalization of the dynamical matrix (Eq. \ref{eq:DynamicalMatrixDef}) is done numerically. This is true also for the QE and ShengBTE program packages and probably results in minor numerical inaccuracies in the calculation of the quantities listed in Tables III and IV. Some phonon labeling could be carried out by continuity arguments for the phonon eigenvectors as a function of wave vector in a particular direction when using relatively dense $\vec{q}$-meshes, but in the absence of a rigorous general approach, the labeling of the phonon modes is left to the default algorithm in QE and ShengBTE.

In the lattice thermal conductivity calculations, both isotopic and three-phonon scattering were included. In the calculation of the isotopic scattering rates, the default Pearson deviation coefficients incorporated in ShengBTE were applied. The so-called proportionality constant scalebroad was set to 0.5 in all ShengBTE calculations (the constant is related to the adaptive Gaussian broadening technique used in ShengBTE for obtaining energy-conserving three-phonon scattering processes).\cite{Li-Gaussian_PhysRevB.85.195436-2012,Li-shengbte-2014} For every structure, all third order atomic force constants were calculated up to 6th-nearest neigbours using the program thirdorder.py included in the ShengBTE distribution.\cite{Li-ThirdOrderPy_PhysRevB.86.174307-2012} The supercells used to calculate the third order atomic force constants with thirdorder.py were $\left(4,4,4\right)$, $\left(3,3,3\right)$ and $\left(2,2,2\right)$ for stuctures \textit{d}-Si, VIII and II, respectively. In all single-point total energy calculations with supercells, only $\Gamma$-point $\vec{k}$-sampling was used and the total energy convergence threshold for self-consistency was set to 10$^{-12}$ a.u.

\section{Results and discussion}
\label{cha:ResultsAndDiscussion}

\subsection{Phonon Spectra}
\label{PhononSpectra}
In the case of \textit{d}-Si, experimental results for phonon eigenvalues $\left\{\omega_{j}\left(\vec{q}\right)\right\}$ are available. \cite{Nilsson-PhysRevB.6.3777_dispersion_exp_Si_alpha-1972} The maximum difference between the experimental values and the eigenvalues obtained in this work is approximately 11\%. Similar computational results have been obtained earlier by using both Local Density Approximation (LDA) and GGA with the plane-wave pseudopotential method \cite{Yin-abInitioPhononsSi-1982,Giannozzi-abInitioPhononsSiGe-1991,Wei-abInitioPhononsSi-1992} \cite{Favot-abInitioPhononsSiGGA-1999}. The maximum difference between the eigenvalues obtained here and the eigenvalues obtained by using LDA with Martins-Troullier-type norm-conserving pseudopotentials is approximately 10\%. \cite{Harkonen-NTE-2014}

\begin{figure}
\includegraphics[width=0.47\textwidth]{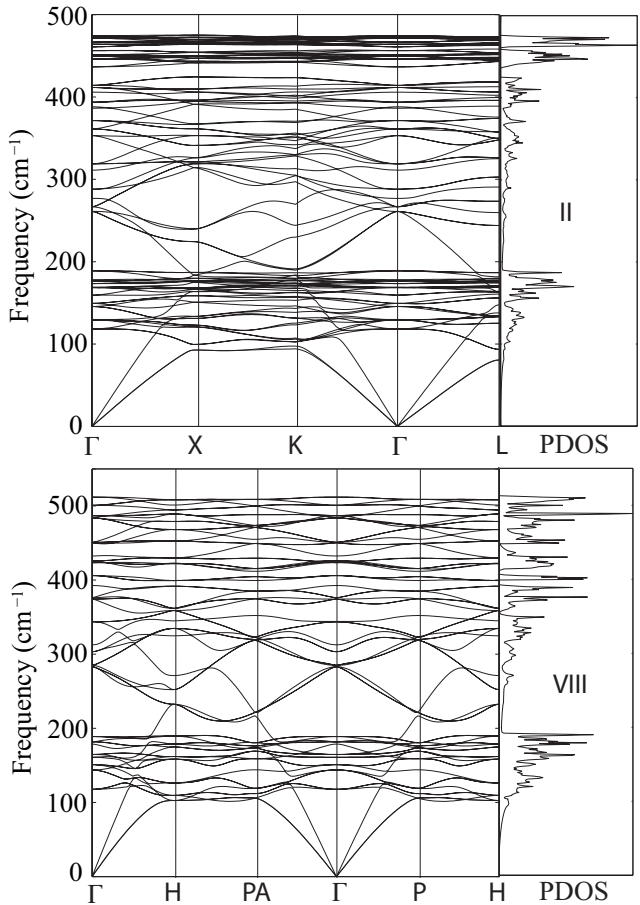}
\caption{Phonon dispersion relations along high symmetry paths in the first Brillouin zone for the clathrate structures II and VIII. PDOS is the phonon density-of-states plot.}
\label{fig:Disp-II-VIII}
\end{figure}

Figure \ref{fig:Disp-II-VIII} shows the calculated phonon dispersion relations for the clathrate structures II and VIII. As in the case of \textit{d}-Si, similar results are obtained by using the present computational methods or LDA with Martins-Troullier-type norm-conserving pseudopotentials. \cite{Harkonen-NTE-2014}
From the similar phonon dispersion relations of the clathrate structures II and VIII it follows that their thermal properties within the harmonic approximation are rather similar, as well. Furthermore, the calculation of anharmonic properties such as thermal expansion and Gr\"{u}neisen parameters within the so-called quasiharmonic approximation have previously lead into similar results for the structures II and VIII. \cite{Harkonen-NTE-2014}

\subsection{Thermal conductivity results}
\label{ThermalConductivityResults}

\begin{figure}
\includegraphics[width=0.47\textwidth]{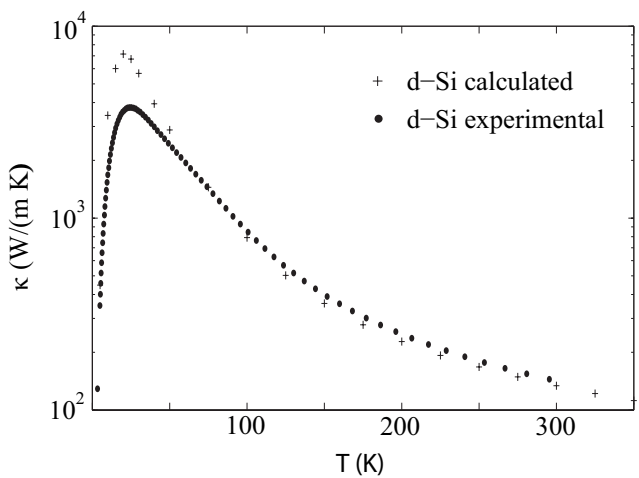}
\caption{The calculated lattice thermal conductivity of \textit{d}-Si and experimental values with natural isotopic composition of Si. \cite{Inyushkin-dSiThermalConductivity-2002}}
\label{fig:d-SiThermalCondExperimental}
\end{figure}

To test the computational methods used in this work, the calculated thermal conductivity results for \textit{d}-Si are compared with the experimental results \cite{Inyushkin-dSiThermalConductivity-2002} in Fig. \ref{fig:d-SiThermalCondExperimental}. 
The maximum difference between the calculational and experimental results is approximately 7-51\% within the temperature range 5-100 K. At temperatures higher than 100 K, the maximum difference between the computational and experimental results is between 4-13\% (the largest difference is at 125 K). The computational results are similar to those obtained earlier for \textit{d}-Si by using DFT-LDA, \cite{Ward-PhysRevB.80.125203-Tcond-2009} or by combining DFT calculation of the force constants and the Green-Kubo formalism applied by using molecular dynamical simulations. \cite{Esfarjani-HeatTransportInSilicon-PhysRevB.84.085204-2011} Acoustic phonon modes have the largest contribution to the lattice thermal conductivity, contributing approximately 100-94\% at temperatures 5-350 K. In particular, the longitudinal acoustic (LA) mode contributes around 0.3-9.4\% at 5-20 K, while within the temperature range 40-350 K its contribution is 22-29\% and the corresponding the two other acoustic modes both contribute about 39-32\%. The contributions of the different phonon modes to lattice thermal conductivity are shown in Table \ref{tab:CalcResultsTConductivity}. These results are in line with the previous results for mode contributions as a function of temperature obtained by using different method. \cite{Esfarjani-HeatTransportInSilicon-PhysRevB.84.085204-2011}

The lattice thermal conductivities calculated here do not include the contributions arising from boundary scattering, which is expected to play a role in the low-temperature regime, but requires empirical, sample-dependent parametrization. \cite{Fugallo-PhysRevB.88.045430-Thermal-Cond-2013} At 100 K, the experimental value is approximately 2 times smaller than the calculated one. The largest MFP at 100 K is approximately $10^{-8}$ m, while the minimum crystal dimension is approximately 2 mm, for which the experimental lattice thermal conductivity values are obtained (similar to those presented in Fig. \ref{fig:d-SiThermalCondExperimental}). \cite{Inyushkin-TCondSilicon-2004} The difference between the experimental and computational results at low temperatures may not be solely explained with the neglection of boundary scattering. One possible reason for this difference is the model used to describe the isotopic scattering. In ShengBTE, the form of isotopic scattering rate is derived in such a way that it is valid for relatively weak perturbations and for rather long wavelengths. \cite{Carruthers-ThermalConductivityLowTemp-1961} Due to this, for example, there might be inaccuracies related to the calculated isotopic scattering rates, making the predicted lattice thermal conductivities for lower temperatures more vulnerable to errors. For higher temperatures, the effect of isotopic scattering decreases. At 350 K, the isotopic scattering decreases the lattice thermal conductivity approximately 2.2\%, while at 100 K the effect is approximately 24.8\%. To test the convergence with respect to the size of the $\vec{q}$-mesh at low temperatures, the lattice thermal conductivity calculation for \textit{d}-Si was also done by using $\left(32,32,32\right)$ $\vec{q}$-mesh at 20 K. The lattice thermal conductivity value obtained was approximately 48\% higher than obtained with $\left(24,24,24\right)$ $\vec{q}$-mesh (at 300 K the value is 2\%), which indicates rather poor convergence at relatively low temperatures. 
\begin{table}
\caption{The contributions of different phonon modes to the total lattice thermal conductivity within two different temperature ranges (all values are in percentages): $T=$ 5-75 K for subscript $l$ and $T=$ 100-350 K for subscript $h$. LA =  longitudinal acoustic mode, 2TA = two transverse acoustic modes, and OP = optical modes.}
		\begin{tabular}{ccccccc}
		\hline\hline
			  Structure & LA$_{l}$\       & 2TA$_{l}$\          & OP$_{l}$\          & LA$_{h}$\       & 2TA$_{h}$\       & OP$_{h}$        \\ \hline
	 \textit{d}-Si  &      0-9        &    100-91           &      0-0           &   22-29         &      78-65       &       0-6       \\
				    II    &      0-26       &    100-65           &      0-8           &   27-25         &      60-51       &       13-24     \\
				  VIII    &      0-22       &    100-74           &      0-4           &   24-26         &      69-58       &       7-16     \\
					\hline\hline
		\end{tabular}
		\label{tab:CalcResultsTConductivity}
\end{table}

\begin{figure}
\includegraphics[width=0.47\textwidth]{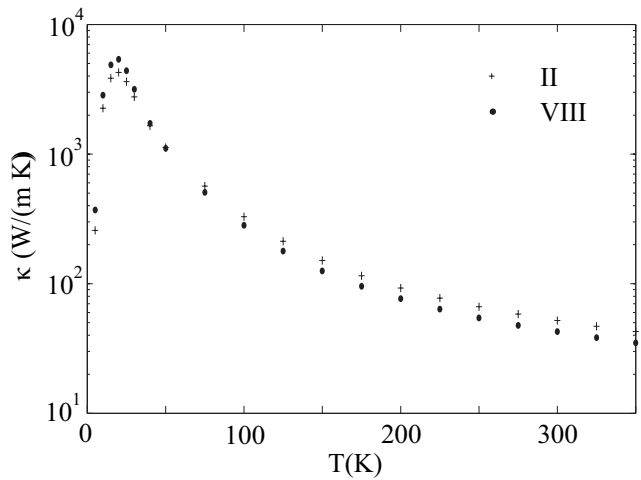}
\caption{The calculated lattice thermal conductivity of the clathrate structures II and VIII.}
\label{fig:ThermalCond_II_VIII}
\end{figure}

The calculated lattice thermal conductivities for the clathrate structures II and VIII are shown in Fig. \ref{fig:ThermalCond_II_VIII}.
The clathrate structure II has lower lattice thermal conductivity than the structure VIII at temperatures below 50 K, while above this temperature the result is opposite. At 300 K, the lattice thermal conductivity values obtained for the clathrate structures II and VIII are approximately 52 and 43 W/(m K), respectively. The results are summarized in Table \ref{tab:CalcResultsTConductivity}. Within the temperature range 100-350 K, the lattice thermal conductivity of the structure VIII is approximately 86-82\% of the values of the structure II. Compared with \textit{d}-Si, the lattice thermal conductivity values of structure II are about 42-38\% of the values of \textit{d}-Si within the temperature range 100-350 K. For the structure VIII, the predicted lattice temperature conductivities are 36-31\% of the values predicted for \textit{d}-Si within the temperature range 100-350 K. In all structures considered, the iterative solution of the BTE converged in 10 steps or less. This indicates that the SMRT solution is rather similar to the iterative solution for these structures. The difference between the SMRT and iterative solution for the lattice thermal conductivity at 300 K is approximately 0.6\% for the clathrate structure II and about 0.4\% for the clathrate structure VIII.

\begin{figure*}
\includegraphics[width=0.99\textwidth]{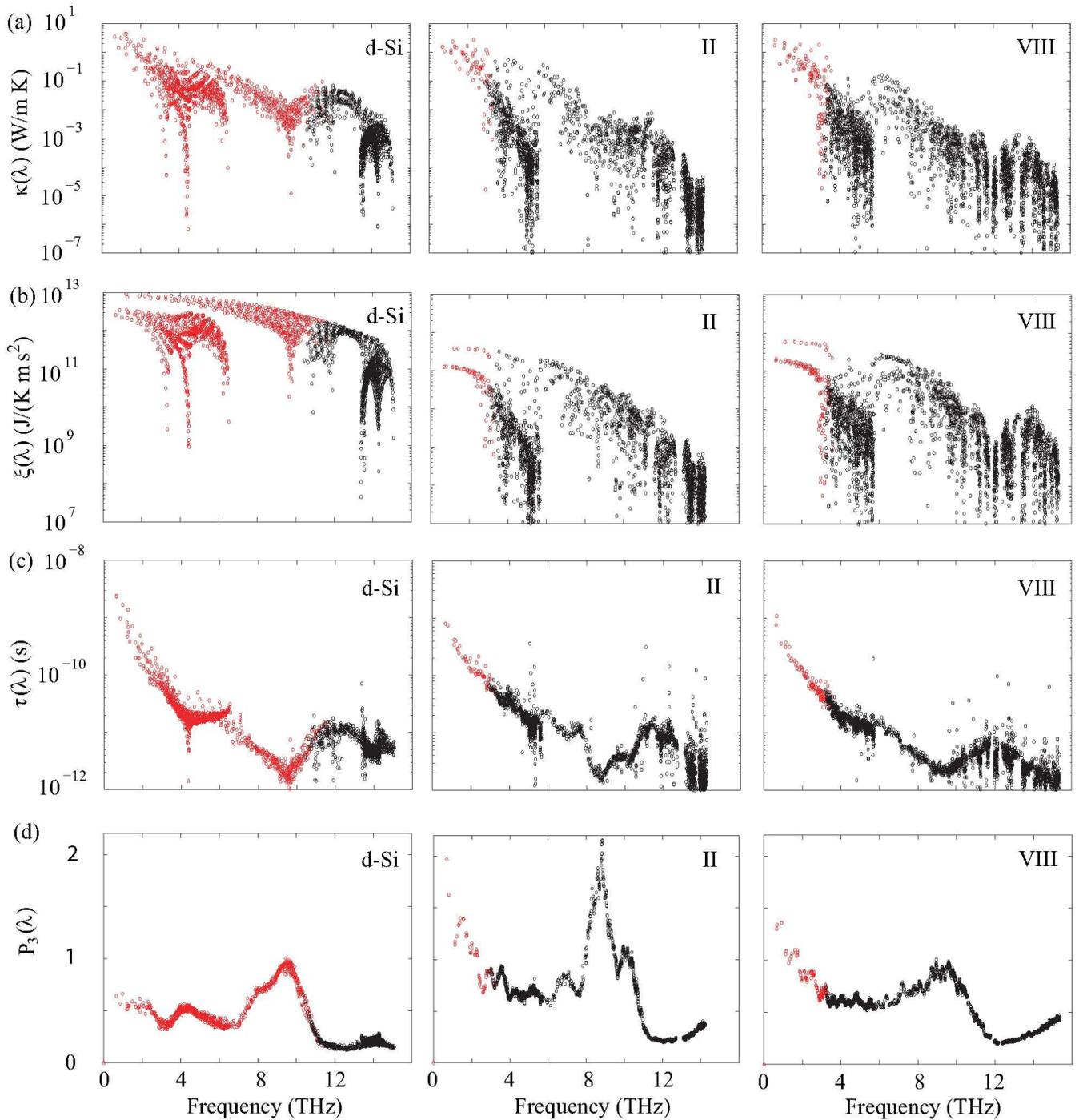}
\caption{The calculated values for each state $\lambda$ as a function of phonon frequency for \textit{d}-Si and the clathrate structures II and VIII at 300 K. (a) The lattice thermal conductivity $\kappa\left(\lambda\right) = 1/3\sum_{\alpha}\kappa_{\alpha}\left(\lambda\right)$ (negative values are not shown), (b) quantities $\xi\left(\lambda\right)$ (Eq. \ref{eq:HeatCGroupVelProdEq_2}), (c) relaxation times $\tau\left(\lambda\right)$ (Eq. \ref{eq:RelaxationTime_modeEq_2}, negative values are not shown) and (d) phonon phase space $P_{3}\left(\lambda\right)$. For all quantities, the acoustic modes are drawn in red and the optical modes in black. The reported $P_{3}\left(\lambda\right)$ values are unitless relative values obtained from $P_{3}\left(\lambda\right)/\max \left\{ \left.P_{3}\left(\lambda\right)\right|_{\textit{d}-Si}\right\}$. (Color online)}
\label{fig:KappaTauXi}
\end{figure*}

The lattice thermal conductivity values for each state as a function of phonon frequency at 300 K are represented in Fig. \ref{fig:KappaTauXi}. In particular, for the clathrate structures II and VIII, the number of acoustic modes within the range 10-0.1 W/m K is to some extent lower than in the case of \textit{d}-Si. The fact that the clathrate structure II (34 atoms in the primitive cell) has higher lattice thermal conductivity than the clathrate structure VIII (23 atoms in the primitive cell) is somewhat unexpected: when comparing structures with similar tetrahedral coordination of the Si atoms, it is usually expected that a structure with the larger unit cell would have lower lattice thermal conductivity. An extreme example of such allotrope of Si with relatively low thermal conductivity values is amorphous Si.

The lattice thermal conductivities calculated here for the structure II are over one order of magnitude higher than the experimental values obtained for hot-pressed pellets prepared from polycrystalline powder samples. \cite{Nolas-ThermalCondSi136-2003} However, the low lattice thermal conductivity values observed in Ref. \cite{Nolas-ThermalCondSi136-2003} have also been suggested to arise from the porosity of the polycrystalline samples. \cite{Shevelkov-AnomalouslyLowTCond-2011} For a better comparison, the experimental values for the lattice thermal conductivity of single crystal samples should be obtained.

\subsection{Group velocity and heat capacity}
\label{GroupVelocityAndHeatCapacity}
In this section, the following term
\begin{equation}
\xi\left(j\right) \equiv \frac{1}{3 V} \sum_{\vec{q},\alpha}  v^{2}_{\alpha}\left(\vec{q}j\right) c_{v}\left(\vec{q}j\right),
\label{eq:HeatCGroupVelProd}
\end{equation}
included in Eq. \ref{eq:DiagonalThermalConductivityCubic} is considered. In Table \ref{tab:CalcResultsHeatCGroupVelProd}, results calculated by using Eq. \ref{eq:HeatCGroupVelProd} for different structures are listed at 300 K. The results in Table \ref{tab:CalcResultsHeatCGroupVelProd} (and \ref{tab:CalcResultsRelaxationTimes}) are shown for the phonon modes 1-6. The modes 1-3 are discussed here as acoustic modes, but we note that due to band crossings with the optical modes the labeling of the modes is not fully rigorous (see computational details).
\begin{table}
\caption{The quantities given by Eq. \ref{eq:HeatCGroupVelProd} at 300 K. Here, units of $\xi\left(j\right)$ are in W/(m K s) $\times 10^{12}$. Only modes 1-6 are considered.}
		\begin{tabular}{ccccccc}
		\hline\hline
			 Structure  &  $\xi\left(1\right)$\ & $\xi\left(2\right)$\ & $\xi\left(3\right)$\ & $\xi\left(4\right)$\ & $\xi\left(5\right)$\ & $\xi\left(6\right)$ \\ \hline
	 \textit{d}-Si  &          0.69         &         1.24         &        2.13          &        0.71          &        0.07          &        0.07  \\
				    II    &          0.07         &         0.08         &        0.08          &        0.02          &        0.02          &        0.02 \\
				  VIII    &          0.09         &         0.10         &        0.11          &        0.01          &        0.01          &        0.02 \\
					\hline\hline
		\end{tabular}
		\label{tab:CalcResultsHeatCGroupVelProd}
\end{table}
The values for \textit{d}-Si are 10-100 times larger than for the clathrate structures. For the clathrate structures II and VIII, the differences are relatively small, indicating that the differences in the lattice thermal conductivity are mainly due to other reasons. In all structures, the optical modes have approximately 10 times smaller values than the acoustic modes, except in the case of \textit{d}-Si, where one optical mode has value in the same order as acoustic modes.

The results listed in Table \ref{tab:CalcResultsHeatCGroupVelProd} manifest the similarity of the harmonic phonon spectra in the clathrate structures II and VIII and point out that to explain their different lattice thermal conductivities, one may find larger differences through the study of the anharmonic interatomic forces and quantities derived from them.

In Fig. \ref{fig:KappaTauXi} the following quantities
\begin{equation}
\xi\left(\lambda\right) \equiv \frac{1}{3 V} \sum_{\alpha}  v^{2}_{\alpha}\left(\lambda\right) c_{v}\left(\lambda\right),
\label{eq:HeatCGroupVelProdEq_2}
\end{equation}
are represented as a function of the phonon frequency. Essentially, the same features can be seen as for $\xi\left(j\right)$ in Table \ref{tab:CalcResultsHeatCGroupVelProd}, but in more detail. In particular, the clathrate structures II and VIII have a similar distribution of the $\xi\left(\lambda\right)$ values, which implies that the largest differences in lattice thermal conductivity are due to the RTs and quantities used to calculate RTs.

\subsection{Phonon lifetimes}
\label{PhononLifetimes}

If the finite lifetime of a phonon mode is assumed to result from the third order atomic force constants, it has been shown that the one-phonon coherent neutron scattering cross section has approximately Lorenzian line shape, provided that the third order atomic forces are relatively weak. \cite{Maradudin-Fein-ScatteringOfNeutrons-1962} The criterion for this is usually taken to be such that $2 \pi \Gamma\left(\vec{q}j\right) / \omega_{j}\left(\vec{q}\right) \ll 1$, where the width of a Lorenzian peak at half maximum $\Gamma\left(\vec{q}j\right)$ of the one-phonon coherent neutron scattering cross section line is related to the relaxation time (or lifetime) of a state $\vec{q}j$ as $\tau\left(\vec{q}j\right) = 1/2\Gamma\left(\vec{q}j\right)$.

To study the validity of SMRT, the percentage of $\vec{q}$-points violating the criterion $2 \pi \Gamma\left(\vec{q}j\right) / \omega_{j}\left(\vec{q}\right) < 0.1$ has been calculated here. In the case of \textit{d}-Si, approximately 1.9\% of the acoustic and 2.5\% of the optical modes violate the criterion at 300 K (mean values for acoustic and optical modes). In the case of the clathrate structure II, the corresponding percentages are 14\% for the acoustic and 52\% for the optical modes. Lastly, in the case of the clathrate structure VIII, the percentages are 6\% for the acoustic and 38\% for the optical modes. For a mode violating this particular criterion, for example with  $\omega_{j}\left(\vec{q}\right) = \textrm{400 cm}^{-1}$, the width at half maximum of a Lorenzian peak would be equal to or larger than 40 cm$^{-1}$. Instead of this, one probably has peak forms deviating from a Lorenzian form.

Next, the following quantities are considered
\begin{equation}
\tau\left(j\right) \equiv  \frac{1}{3}\sum_{\vec{q},\alpha}  \tau_{\alpha}\left(\vec{q}j\right).
\label{eq:RelaxationTime_mode}
\end{equation}
The results are listed in Table \ref{tab:CalcResultsRelaxationTimes} at 300 K. 
\begin{table}
\caption{The quantities given by Eq. \ref{eq:RelaxationTime_mode} at 300 K. Here, the units of $\tau\left(j\right)$ are in s $\times 10^{-10}$. Only modes 1-6 are considered.}
		\begin{tabular}{ccccccc}
		\hline\hline
			  Structure & $\tau\left(1\right)$\ &$\tau\left(2\right)$\ &$\tau\left(3\right)$\ &$\tau\left(4\right)$\ &$\tau\left(5\right)$\ & $\tau\left(6\right)$ \\ \hline
	 \textit{d}-Si  &          0.37         &         0.28         &        0.08          &        0.08          &        0.06          &        0.06  \\
				    II    &          1.44         &         1.47         &        0.80          &        0.51          &        0.43          &        0.39  \\
				  VIII    &          0.90         &         1.00         &        0.48          &        0.33          &        0.29          &        0.23  \\
					\hline\hline
		\end{tabular}
		\label{tab:CalcResultsRelaxationTimes}
\end{table}
The smallest RTs are obtained for \textit{d}-Si. For acoustic modes, the clathrate structures II and VIII have approximately 3 and 2 times higher values, resprectively. For optical modes considered, the difference is approximately two times larger. For \textit{d}-Si, the percentage of acoustic states for which $\tau\left(\lambda\right)< 0$ at 300 K is between 0-0.5\%, while for the clathrate structures II and VIII these percentages are 0\% and 0-2\%, respectively.

In Fig. \ref{fig:KappaTauXi} the following quantities
\begin{equation}
\tau\left(\lambda\right) \equiv  \frac{1}{3}\sum_{\alpha}  \tau_{\alpha}\left(\lambda\right),
\label{eq:RelaxationTime_modeEq_2}
\end{equation}
are represented as a function of phonon frequency. The results of Fig. \ref{fig:KappaTauXi} appear to contradict with the results given by Table \ref{tab:CalcResultsRelaxationTimes}, since for \textit{d}-Si it seems that the acoustic modes have in general larger values for the lifetimes than in the case of the clathrate structure II. However, in the case of the clathrate structure II, the values for $\tau\left(\lambda\right)$ are clustered within the range $10^{-9}-10^{-10}$ s, while for \textit{d}-Si there are more values of $\tau\left(\lambda\right)$ for acoustic modes within the range $10^{-11}-10^{-12}$ s, as well. The difference between the RTs of the acoustic modes in the clathrate structures II and VIII is rather small but can be identified from Fig. \ref{fig:KappaTauXi}, structure VIII having slightly lower values in general.

To summarize, within the current formalism, the results given by Tables \ref{tab:CalcResultsHeatCGroupVelProd}, \ref{tab:CalcResultsRelaxationTimes} and Fig. \ref{fig:KappaTauXi} indicate that the lower lattice thermal conductivity of the clathrate structure VIII can be partly explained with relatively short RTs and with differing harmonic phonon spectra, that is, $\xi\left(j\right)$ values lower than for \textit{d}-Si and similar to those of structure II.

\subsection{Phonon phase space}
\label{PhononPhaseSpace}

The so-called phonon phase space can be used as a measure of available scattering channels for phonons obeying the conservation of energy and quasi-momentum. The transition probabilities, for example in Eq. \ref{eq:BTEGoldenRule_2}, are proportional to the phonon phase space. The phase space for processes in which one phonon is annihilated $\left(-\right)$ or two phonons are annihilated $\left(+\right)$, can be written as \cite{Lindsay-ThreePhononPhaseSpace-2008,Li-shengbte-2014}
\begin{eqnarray}
&&P^{\pm}_{3}\left(\vec{q}j\right) \equiv \sum_{j',j''} \sum_{\vec{q}'} \nonumber \\
&& \times \delta\left[\omega_{j}\left(\vec{q}\right) \pm \omega_{j'}\left(\vec{q}'\right) - \omega_{j}\left(\vec{q}+\vec{q}'-\vec{G}\right)\right], 
\label{eq:PhaseSpaceEq_1}
\end{eqnarray}
and the total phase space is given as a sum of these terms times some normalization factor (one may replace the summation over $\vec{q}'$ by integral in the continuum limit). In Fig. \ref{fig:KappaTauXi}, the results based on the following relation (times some normalization factor $1/\Omega$)
\begin{equation}
P_{3}\left(\lambda\right) = \frac{1}{\Omega}\left[P^{+}_{3}\left(\lambda\right) + P^{-}_{3}\left(\lambda\right)\right],
\label{eq:PhaseSpaceEq_2}
\end{equation}
are represented. It is usually expected that the larger values of $P_{3}\left(\lambda\right)$ result in smaller values for $\tau\left(\lambda\right)$. However, this is not necessarily the case since the values of $V\left(\lambda;\lambda';\lambda''\right)$ and $\bar{n}_{\lambda}\bar{n}_{\lambda'}\left(\bar{n}_{\lambda''}+1\right)$ have some significance and may change the outcome to opposite. Indeed, it can be seen from Fig. \ref{fig:KappaTauXi} that with frequency values $8-10$ THz, there is some correlation between the relaxation times and phase space values. The largest values of $P_{3}\left(\lambda\right)$ are obtained for the clathrate stucture II, which is true for acoustical and optical modes in general. Due to this observation and since the harmonic phonon spectra is rather similar for the clathrate structures II and VIII, it seems that the shorter RTs of structure VIII are caused by the differences in factors proportional to $\left|V\left(\lambda;\lambda';\lambda''\right)\right|^{2}\bar{n}_{\lambda}\bar{n}_{\lambda'}\left(\bar{n}_{\lambda''}+1\right)$ (see for instance Eqs. \ref{eq:BTETotal_3} and \ref{eq:BTEGoldenRule_2}).

The phase space results for the clathrate structures II and VIII show that seemingly small changes in the harmonic phonon spectra, even for structures with similar thermodynamical properties arising from the harmonic phonon spectra, can lead to relatively large differences in the phase space calculations.

\subsection{Mean free paths}
\label{MeanFreePaths}
One criterion for evaluating the validity of the propagating phonon picture is to calculate the length of the MFP of a phonon state. It has been suggested that the propagating phonon picture is a reasonable approximation to work with if the mean free paths are larger than the distance between the nearest atoms in the crystal. \cite{Allen-ThermalCondGlass-1993} This may cause some problems from a physical point of view since in the present BTE approach, one considers phonons propagating in the lattice with wave lengths larger than the distance between the nearest atoms and on the other hand the obtained results may imply that MFPs are smaller than this distance, while such phonons do not exist in the harmonic spectra. In these cases, the vibrational states of a crystal may be considered as a linear combination of the harmonic states with some periodic time dependence. Classification of different vibrational states, which cannot be described as phonons, has been made. \cite{Allen-DiffusonsLoconsPropagons-1999} For states with MFPs shorter than interatomic distances, the lattice thermal conductivity may not be described by the heat flux given by Eq. \ref{eq:HeatFluxConductivity} in proper detail, but the use of a more general heat flux might yield further information on the mechanisms of lattice thermal conductivity and could change the outcome of the present results up to some degree. Such heat fluxes have been derived \cite{Hardy-PhysRev.132.168-Energy-Flux-1963} and used to calculate lattice thermal conductivity values of amorphous Si by using linear response approach. \cite{Allen-ThermalCondGlass-1993}

The MFPs represented here are calculated by using
\begin{equation}
\textrm{mfp}\left(\lambda\right) \equiv \tau\left(\lambda\right) \left|\vec{v}\left(\lambda\right) \right|, \quad \tau\left(\lambda\right) \equiv \frac{1}{3} \left|\sum_{\alpha}  \tau_{\alpha}\left(\lambda\right) \right|.
\label{eq:MeanFreePathDef}
\end{equation}
For \textit{d}-Si, the percentage of $\vec{q}$-points violating the criterion $\textrm{mfp}\left(\lambda\right) < a$ ($a$ being the lattice constant of the primitive unit cell) is 0.5-1.0\% (the values for acoustic modes are between 0.5-1.0\%) and 0.7-2.4\% (optical modes) at 300 K. For the clathrate structure II, the same percentages are 0\% (acoustic modes), 0-86\% (optical modes), while for the clathrate structure VIII they are 2\% (acoustic modes) and 2-96\% (optical modes). The largest percentages for the clathrate structures II and VIII are about 100 times higher than for \textit{d}-Si. In any case, the largest lattice thermal conductivity contribution arises from the states which do not violate the above MFP criterion. However, one may ask, how to describe the states violating the MFP criterion and what kind of errors might arise from the present perturbative approach where the harmonic states are assumed to be stationary.

\section{Conclusions}
\label{cha:Conclusions}
The lattice thermal conductivity of silicon clathrate frameworks II and VIII was investigated by using \textit{ab initio} DFT lattice dynamics with an iterative solution of the Boltzmann transport equation. At 100-350 K the lattice thermal conductivities of the clathrate structures II and VIII are approximately 42-38\% and 36-31\% of the lattice thermal conductivity of \textit{d}-Si, respectively. While the harmonic phonon spectra of the studied clathrate frameworks are rather similar, the lattice thermal conductivity values of the framework VIII at 100-350 K are approximately 14-18\% lower than for the framework II. Analysis of the results shows that the difference in the lattice thermal conductivity of the clathrate structures II and VIII is due to the shorter relaxation times of the acoustic modes in the structure VIII, which seem to arise in part from the stronger 3-phonon interaction coefficients $V\left(\lambda;\lambda';\lambda''\right)$. In the present approach, the harmonic phonon frequency shifts due to anharmonic interatomic forces are neglected, which is a point of improvement for future studies in this field. In particular, the three-phonon phase space was found to be a rather sensitive for the changes in the harmonic phonon spectra when comparing the clathrate structures II and VIII. Thus, more detailed calculations on the lattice thermal conductivities of the clathrate structures II and VIII could change the results obtained here up to some degree, for instance, by taking into account the harmonic phonon frequency shifts due to the third and fourth order interatomic forces. Overall, the present analysis on the lattice thermal conductivity of the clathrate frameworks II and VIII may facilitate further theoretical work towards more detailed understanding of the lattice thermal conductivity and thermoelectric efficiency of semiconducting clathrate materials.

\begin{acknowledgments}
We gratefully acknowledge funding from the Foundation for Research of Natural Resources in Finland (grant 17591/13). The computing resources were provided by the Finnish Grid Infrastructure (FGI) and CSC - the Finnish IT Center for Science. We also thank Prof. Robert van Leeuwen and Dr. Gerrit Groenhof (University of Jyv\"{a}skyl\"{a}) for useful discussions on various aspects of the present work.
\end{acknowledgments}
\bibliography{bibfile}

\begin{thebibliography}{51}%
\makeatletter
\providecommand \@ifxundefined [1]{%
 \@ifx{#1\undefined}
}%
\providecommand \@ifnum [1]{%
 \ifnum #1\expandafter \@firstoftwo
 \else \expandafter \@secondoftwo
 \fi
}%
\providecommand \@ifx [1]{%
 \ifx #1\expandafter \@firstoftwo
 \else \expandafter \@secondoftwo
 \fi
}%
\providecommand \natexlab [1]{#1}%
\providecommand \enquote  [1]{``#1''}%
\providecommand \bibnamefont  [1]{#1}%
\providecommand \bibfnamefont [1]{#1}%
\providecommand \citenamefont [1]{#1}%
\providecommand \href@noop [0]{\@secondoftwo}%
\providecommand \href [0]{\begingroup \@sanitize@url \@href}%
\providecommand \@href[1]{\@@startlink{#1}\@@href}%
\providecommand \@@href[1]{\endgroup#1\@@endlink}%
\providecommand \@sanitize@url [0]{\catcode `\\12\catcode `\$12\catcode
  `\&12\catcode `\#12\catcode `\^12\catcode `\_12\catcode `\%12\relax}%
\providecommand \@@startlink[1]{}%
\providecommand \@@endlink[0]{}%
\providecommand \url  [0]{\begingroup\@sanitize@url \@url }%
\providecommand \@url [1]{\endgroup\@href {#1}{\urlprefix }}%
\providecommand \urlprefix  [0]{URL }%
\providecommand \Eprint [0]{\href }%
\providecommand \doibase [0]{http://dx.doi.org/}%
\providecommand \selectlanguage [0]{\@gobble}%
\providecommand \bibinfo  [0]{\@secondoftwo}%
\providecommand \bibfield  [0]{\@secondoftwo}%
\providecommand \translation [1]{[#1]}%
\providecommand \BibitemOpen [0]{}%
\providecommand \bibitemStop [0]{}%
\providecommand \bibitemNoStop [0]{.\EOS\space}%
\providecommand \EOS [0]{\spacefactor3000\relax}%
\providecommand \BibitemShut  [1]{\csname bibitem#1\endcsname}%
\let\auto@bib@innerbib\@empty
\bibitem [{\citenamefont {Ioffe}(1957)}]{Ioffe-SemicondThermoelements-1957}%
  \BibitemOpen
  \bibfield  {author} {\bibinfo {author} {\bibfnamefont {A.~F.}\ \bibnamefont
  {Ioffe}},\ }\href@noop {} {\emph {\bibinfo {title} {Semiconductor
  Thermoelements and Thermoelectric Cooling}}}\ (\bibinfo  {publisher}
  {Infosearch Limited London},\ \bibinfo {year} {1957})\ pp.\ \bibinfo {pages}
  {1--184}\BibitemShut {NoStop}%
\bibitem [{\citenamefont {Slack}\ and\ \citenamefont
  {Rowe}(1995)}]{Slack-crtThermoelectrics-1995}%
  \BibitemOpen
  \bibfield  {author} {\bibinfo {author} {\bibfnamefont {G.~A.}\ \bibnamefont
  {Slack}}\ and\ \bibinfo {author} {\bibfnamefont {D.}~\bibnamefont {Rowe}},\
  }\href@noop {} {\bibfield  {journal} {\bibinfo  {journal} {CRC, Boca Raton,
  FL}\ ,\ \bibinfo {pages} {407}} (\bibinfo {year} {1995})}\BibitemShut
  {NoStop}%
\bibitem [{\citenamefont {Takabatake}\ \emph {et~al.}(2014)\citenamefont
  {Takabatake}, \citenamefont {Suekuni}, \citenamefont {Nakayama},\ and\
  \citenamefont {Kaneshita}}]{Takabatake-RevModPhys.86.669-2014}%
  \BibitemOpen
  \bibfield  {author} {\bibinfo {author} {\bibfnamefont {T.}~\bibnamefont
  {Takabatake}}, \bibinfo {author} {\bibfnamefont {K.}~\bibnamefont {Suekuni}},
  \bibinfo {author} {\bibfnamefont {T.}~\bibnamefont {Nakayama}}, \ and\
  \bibinfo {author} {\bibfnamefont {E.}~\bibnamefont {Kaneshita}},\ }\href
  {\doibase 10.1103/RevModPhys.86.669} {\bibfield  {journal} {\bibinfo
  {journal} {Rev. Mod. Phys.}\ }\textbf {\bibinfo {volume} {86}},\ \bibinfo
  {pages} {669} (\bibinfo {year} {2014})}\BibitemShut {NoStop}%
\bibitem [{\citenamefont
  {Callaway}(1959)}]{Callaway-PhysRev.113.1046-LatCond-1959}%
  \BibitemOpen
  \bibfield  {author} {\bibinfo {author} {\bibfnamefont {J.}~\bibnamefont
  {Callaway}},\ }\href {\doibase 10.1103/PhysRev.113.1046} {\bibfield
  {journal} {\bibinfo  {journal} {Phys. Rev.}\ }\textbf {\bibinfo {volume}
  {113}},\ \bibinfo {pages} {1046} (\bibinfo {year} {1959})}\BibitemShut
  {NoStop}%
\bibitem [{\citenamefont
  {Allen}(2013)}]{Allen-PhysRevB.88.144302-Improved-Callaway-2013}%
  \BibitemOpen
  \bibfield  {author} {\bibinfo {author} {\bibfnamefont {P.~B.}\ \bibnamefont
  {Allen}},\ }\href {\doibase 10.1103/PhysRevB.88.144302} {\bibfield  {journal}
  {\bibinfo  {journal} {Phys. Rev. B}\ }\textbf {\bibinfo {volume} {88}},\
  \bibinfo {pages} {144302} (\bibinfo {year} {2013})}\BibitemShut {NoStop}%
\bibitem [{\citenamefont {Omini}\ and\ \citenamefont
  {Sparavigna}(1995)}]{Omini-iterative-BTE-1995}%
  \BibitemOpen
  \bibfield  {author} {\bibinfo {author} {\bibfnamefont {M.}~\bibnamefont
  {Omini}}\ and\ \bibinfo {author} {\bibfnamefont {A.}~\bibnamefont
  {Sparavigna}},\ }\href@noop {} {\bibfield  {journal} {\bibinfo  {journal}
  {Physica B: Condensed Matter}\ }\textbf {\bibinfo {volume} {212}},\ \bibinfo
  {pages} {101} (\bibinfo {year} {1995})}\BibitemShut {NoStop}%
\bibitem [{\citenamefont {Omini}\ and\ \citenamefont
  {Sparavigna}(1996)}]{Omini-PhysRevB.53.9064-iterative-BTE-1996}%
  \BibitemOpen
  \bibfield  {author} {\bibinfo {author} {\bibfnamefont {M.}~\bibnamefont
  {Omini}}\ and\ \bibinfo {author} {\bibfnamefont {A.}~\bibnamefont
  {Sparavigna}},\ }\href {\doibase 10.1103/PhysRevB.53.9064} {\bibfield
  {journal} {\bibinfo  {journal} {Phys. Rev. B}\ }\textbf {\bibinfo {volume}
  {53}},\ \bibinfo {pages} {9064} (\bibinfo {year} {1996})}\BibitemShut
  {NoStop}%
\bibitem [{\citenamefont {Broido}\ \emph {et~al.}(2005)\citenamefont {Broido},
  \citenamefont {Ward},\ and\ \citenamefont
  {Mingo}}]{Broido-PhysRevB.72.014308-LatCond-BTE-2005}%
  \BibitemOpen
  \bibfield  {author} {\bibinfo {author} {\bibfnamefont {D.~A.}\ \bibnamefont
  {Broido}}, \bibinfo {author} {\bibfnamefont {A.}~\bibnamefont {Ward}}, \ and\
  \bibinfo {author} {\bibfnamefont {N.}~\bibnamefont {Mingo}},\ }\href
  {\doibase 10.1103/PhysRevB.72.014308} {\bibfield  {journal} {\bibinfo
  {journal} {Phys. Rev. B}\ }\textbf {\bibinfo {volume} {72}},\ \bibinfo
  {pages} {014308} (\bibinfo {year} {2005})}\BibitemShut {NoStop}%
\bibitem [{\citenamefont {Fugallo}\ \emph {et~al.}(2013)\citenamefont
  {Fugallo}, \citenamefont {Lazzeri}, \citenamefont {Paulatto},\ and\
  \citenamefont {Mauri}}]{Fugallo-PhysRevB.88.045430-Thermal-Cond-2013}%
  \BibitemOpen
  \bibfield  {author} {\bibinfo {author} {\bibfnamefont {G.}~\bibnamefont
  {Fugallo}}, \bibinfo {author} {\bibfnamefont {M.}~\bibnamefont {Lazzeri}},
  \bibinfo {author} {\bibfnamefont {L.}~\bibnamefont {Paulatto}}, \ and\
  \bibinfo {author} {\bibfnamefont {F.}~\bibnamefont {Mauri}},\ }\href
  {\doibase 10.1103/PhysRevB.88.045430} {\bibfield  {journal} {\bibinfo
  {journal} {Phys. Rev. B}\ }\textbf {\bibinfo {volume} {88}},\ \bibinfo
  {pages} {045430} (\bibinfo {year} {2013})}\BibitemShut {NoStop}%
\bibitem [{\citenamefont {Li}\ \emph {et~al.}(2014)\citenamefont {Li},
  \citenamefont {Carrete}, \citenamefont {Katcho},\ and\ \citenamefont
  {Mingo}}]{Li-shengbte-2014}%
  \BibitemOpen
  \bibfield  {author} {\bibinfo {author} {\bibfnamefont {W.}~\bibnamefont
  {Li}}, \bibinfo {author} {\bibfnamefont {J.}~\bibnamefont {Carrete}},
  \bibinfo {author} {\bibfnamefont {N.~A.}\ \bibnamefont {Katcho}}, \ and\
  \bibinfo {author} {\bibfnamefont {N.}~\bibnamefont {Mingo}},\ }\href@noop {}
  {\bibfield  {journal} {\bibinfo  {journal} {Comput. Phys. Commun.}\ }\textbf
  {\bibinfo {volume} {185}},\ \bibinfo {pages} {1747} (\bibinfo {year}
  {2014})}\BibitemShut {NoStop}%
\bibitem [{\citenamefont {Chernatynskiy}\ and\ \citenamefont
  {Phillpot}(2015)}]{Chernatynskiy-PhonTS-2015}%
  \BibitemOpen
  \bibfield  {author} {\bibinfo {author} {\bibfnamefont {A.}~\bibnamefont
  {Chernatynskiy}}\ and\ \bibinfo {author} {\bibfnamefont {S.~R.}\ \bibnamefont
  {Phillpot}},\ }\href@noop {} {\bibfield  {journal} {\bibinfo  {journal}
  {Comput. Phys. Commun.}\ }\textbf {\bibinfo {volume} {192}},\ \bibinfo
  {pages} {196} (\bibinfo {year} {2015})}\BibitemShut {NoStop}%
\bibitem [{\citenamefont {Kasper}\ \emph {et~al.}(1965)\citenamefont {Kasper},
  \citenamefont {Hagenmuller}, \citenamefont {Pouchard},\ and\ \citenamefont
  {Cros}}]{Kasper-24121965.science.clath.1965}%
  \BibitemOpen
  \bibfield  {author} {\bibinfo {author} {\bibfnamefont {J.~S.}\ \bibnamefont
  {Kasper}}, \bibinfo {author} {\bibfnamefont {P.}~\bibnamefont {Hagenmuller}},
  \bibinfo {author} {\bibfnamefont {M.}~\bibnamefont {Pouchard}}, \ and\
  \bibinfo {author} {\bibfnamefont {C.}~\bibnamefont {Cros}},\ }\href {\doibase
  10.1126/science.150.3704.1713} {\bibfield  {journal} {\bibinfo  {journal}
  {Science}\ }\textbf {\bibinfo {volume} {150}},\ \bibinfo {pages} {1713}
  (\bibinfo {year} {1965})}\BibitemShut {NoStop}%
\bibitem [{\citenamefont {Sootsman}\ \emph {et~al.}(2009)\citenamefont
  {Sootsman}, \citenamefont {Chung},\ and\ \citenamefont
  {Kanatzidis}}]{Sootsman-thermoelect.-review.2009-ANIE200900598}%
  \BibitemOpen
  \bibfield  {author} {\bibinfo {author} {\bibfnamefont {J.}~\bibnamefont
  {Sootsman}}, \bibinfo {author} {\bibfnamefont {D.}~\bibnamefont {Chung}}, \
  and\ \bibinfo {author} {\bibfnamefont {M.}~\bibnamefont {Kanatzidis}},\
  }\href {\doibase 10.1002/anie.200900598} {\bibfield  {journal} {\bibinfo
  {journal} {Angew. Chem. Int. Ed.}\ }\textbf {\bibinfo {volume} {48}},\
  \bibinfo {pages} {8616} (\bibinfo {year} {2009})}\BibitemShut {NoStop}%
\bibitem [{\citenamefont {Christensen}\ \emph {et~al.}(2010)\citenamefont
  {Christensen}, \citenamefont {Johnsen},\ and\ \citenamefont
  {Iversen}}]{Christensen.et.al.-2010-B916400F}%
  \BibitemOpen
  \bibfield  {author} {\bibinfo {author} {\bibfnamefont {M.}~\bibnamefont
  {Christensen}}, \bibinfo {author} {\bibfnamefont {S.}~\bibnamefont
  {Johnsen}}, \ and\ \bibinfo {author} {\bibfnamefont {B.~B.}\ \bibnamefont
  {Iversen}},\ }\href {\doibase 10.1039/B916400F} {\bibfield  {journal}
  {\bibinfo  {journal} {Dalton Trans.}\ }\textbf {\bibinfo {volume} {39}},\
  \bibinfo {pages} {978} (\bibinfo {year} {2010})}\BibitemShut {NoStop}%
\bibitem [{\citenamefont {Nolas}\ \emph {et~al.}(1998)\citenamefont {Nolas},
  \citenamefont {Cohn}, \citenamefont {Slack},\ and\ \citenamefont
  {Schujman}}]{Nolas-Ge-clath-thermoelect-1998}%
  \BibitemOpen
  \bibfield  {author} {\bibinfo {author} {\bibfnamefont {G.}~\bibnamefont
  {Nolas}}, \bibinfo {author} {\bibfnamefont {J.}~\bibnamefont {Cohn}},
  \bibinfo {author} {\bibfnamefont {G.}~\bibnamefont {Slack}}, \ and\ \bibinfo
  {author} {\bibfnamefont {S.}~\bibnamefont {Schujman}},\ }\href {\doibase
  10.1063/1.121747} {\bibfield  {journal} {\bibinfo  {journal} {Applied Physics
  Letters}\ }\textbf {\bibinfo {volume} {73}},\ \bibinfo {pages} {178}
  (\bibinfo {year} {1998})}\BibitemShut {NoStop}%
\bibitem [{\citenamefont {Cohn}\ \emph {et~al.}(1999)\citenamefont {Cohn},
  \citenamefont {Nolas}, \citenamefont {Fessatidis}, \citenamefont {Metcalf},\
  and\ \citenamefont {Slack}}]{Cohn-GlasslikeHeatCond-PhysRevLett.82.779-1999}%
  \BibitemOpen
  \bibfield  {author} {\bibinfo {author} {\bibfnamefont {J.~L.}\ \bibnamefont
  {Cohn}}, \bibinfo {author} {\bibfnamefont {G.~S.}\ \bibnamefont {Nolas}},
  \bibinfo {author} {\bibfnamefont {V.}~\bibnamefont {Fessatidis}}, \bibinfo
  {author} {\bibfnamefont {T.~H.}\ \bibnamefont {Metcalf}}, \ and\ \bibinfo
  {author} {\bibfnamefont {G.~A.}\ \bibnamefont {Slack}},\ }\href {\doibase
  10.1103/PhysRevLett.82.779} {\bibfield  {journal} {\bibinfo  {journal} {Phys.
  Rev. Lett.}\ }\textbf {\bibinfo {volume} {82}},\ \bibinfo {pages} {779}
  (\bibinfo {year} {1999})}\BibitemShut {NoStop}%
\bibitem [{\citenamefont {Tadano}\ \emph {et~al.}(2015)\citenamefont {Tadano},
  \citenamefont {Gohda},\ and\ \citenamefont
  {Tsuneyuki}}]{Tadano-ImpactOfRattlers-PhysRevLett.114.095501-2015}%
  \BibitemOpen
  \bibfield  {author} {\bibinfo {author} {\bibfnamefont {T.}~\bibnamefont
  {Tadano}}, \bibinfo {author} {\bibfnamefont {Y.}~\bibnamefont {Gohda}}, \
  and\ \bibinfo {author} {\bibfnamefont {S.}~\bibnamefont {Tsuneyuki}},\ }\href
  {\doibase 10.1103/PhysRevLett.114.095501} {\bibfield  {journal} {\bibinfo
  {journal} {Phys. Rev. Lett.}\ }\textbf {\bibinfo {volume} {114}},\ \bibinfo
  {pages} {095501} (\bibinfo {year} {2015})}\BibitemShut {NoStop}%
\bibitem [{\citenamefont {Nolas}\ \emph {et~al.}(2003)\citenamefont {Nolas},
  \citenamefont {Beekman}, \citenamefont {Gryko}, \citenamefont {Lamberton~Jr},
  \citenamefont {Tritt},\ and\ \citenamefont
  {McMillan}}]{Nolas-ThermalCondSi136-2003}%
  \BibitemOpen
  \bibfield  {author} {\bibinfo {author} {\bibfnamefont {G.}~\bibnamefont
  {Nolas}}, \bibinfo {author} {\bibfnamefont {M.}~\bibnamefont {Beekman}},
  \bibinfo {author} {\bibfnamefont {J.}~\bibnamefont {Gryko}}, \bibinfo
  {author} {\bibfnamefont {G.}~\bibnamefont {Lamberton~Jr}}, \bibinfo {author}
  {\bibfnamefont {T.}~\bibnamefont {Tritt}}, \ and\ \bibinfo {author}
  {\bibfnamefont {P.}~\bibnamefont {McMillan}},\ }\href@noop {} {\bibfield
  {journal} {\bibinfo  {journal} {Appl. Phys. lett.}\ }\textbf {\bibinfo
  {volume} {82}},\ \bibinfo {pages} {910} (\bibinfo {year} {2003})}\BibitemShut
  {NoStop}%
\bibitem [{\citenamefont {Huang}\ and\ \citenamefont
  {Born}(1954)}]{Born-Huang-DynamicalTheory-1954}%
  \BibitemOpen
  \bibfield  {author} {\bibinfo {author} {\bibfnamefont {K.}~\bibnamefont
  {Huang}}\ and\ \bibinfo {author} {\bibfnamefont {M.}~\bibnamefont {Born}},\
  }\href@noop {} {\emph {\bibinfo {title} {Dynamical Theory of Crystal
  Lattices}}}\ (\bibinfo  {publisher} {Clarendon Press Oxford},\ \bibinfo
  {year} {1954})\ pp.\ \bibinfo {pages} {293--306}\BibitemShut {NoStop}%
\bibitem [{\citenamefont {Maradudin}\ \emph {et~al.}(1971)\citenamefont
  {Maradudin}, \citenamefont {Montroll}, \citenamefont {Weiss},\ and\
  \citenamefont {Ipatova}}]{Maradudin-harm-appr-1971}%
  \BibitemOpen
  \bibfield  {author} {\bibinfo {author} {\bibfnamefont {A.}~\bibnamefont
  {Maradudin}}, \bibinfo {author} {\bibfnamefont {E.}~\bibnamefont {Montroll}},
  \bibinfo {author} {\bibfnamefont {G.}~\bibnamefont {Weiss}}, \ and\ \bibinfo
  {author} {\bibfnamefont {I.}~\bibnamefont {Ipatova}},\ }\href@noop {} {\emph
  {\bibinfo {title} {Theory of The Lattice Dynamics in The Harmonic
  Approximation}}},\ Vol.\ \bibinfo {volume} {Supplement 3}\ (\bibinfo
  {publisher} {Academic Press},\ \bibinfo {year} {1971})\ pp.\ \bibinfo {pages}
  {6--57}\BibitemShut {NoStop}%
\bibitem [{\citenamefont {Maradudin}(1974)}]{Maradudin-dyn-prop-solids-1974}%
  \BibitemOpen
  \bibfield  {author} {\bibinfo {author} {\bibfnamefont {A.}~\bibnamefont
  {Maradudin}},\ }\href@noop {} {\emph {\bibinfo {title} {Elements of The
  Theory of Lattice Dynamics}}},\ Vol.~\bibinfo {volume} {1}\ (\bibinfo
  {publisher} {North-Holland Publishing Company},\ \bibinfo {year} {1974})\
  pp.\ \bibinfo {pages} {1--82}\BibitemShut {NoStop}%
\bibitem [{\citenamefont {Leibfried}\ \emph {et~al.}(1961)\citenamefont
  {Leibfried}, \citenamefont {Ludwig}, \citenamefont {Seitz},\ and\
  \citenamefont {Turnbull}}]{Leibfried-SolidStatePhysics-1961}%
  \BibitemOpen
  \bibfield  {author} {\bibinfo {author} {\bibfnamefont {G.}~\bibnamefont
  {Leibfried}}, \bibinfo {author} {\bibfnamefont {W.}~\bibnamefont {Ludwig}},
  \bibinfo {author} {\bibfnamefont {F.}~\bibnamefont {Seitz}}, \ and\ \bibinfo
  {author} {\bibfnamefont {D.}~\bibnamefont {Turnbull}},\ }\href@noop {}
  {\bibfield  {journal} {\bibinfo  {journal} {Academic Press, New York}\
  }\textbf {\bibinfo {volume} {12}},\ \bibinfo {pages} {275} (\bibinfo {year}
  {1961})}\BibitemShut {NoStop}%
\bibitem [{\citenamefont
  {Hardy}(1963)}]{Hardy-PhysRev.132.168-Energy-Flux-1963}%
  \BibitemOpen
  \bibfield  {author} {\bibinfo {author} {\bibfnamefont {R.~J.}\ \bibnamefont
  {Hardy}},\ }\href {\doibase 10.1103/PhysRev.132.168} {\bibfield  {journal}
  {\bibinfo  {journal} {Phys. Rev.}\ }\textbf {\bibinfo {volume} {132}},\
  \bibinfo {pages} {168} (\bibinfo {year} {1963})}\BibitemShut {NoStop}%
\bibitem [{\citenamefont {Ziman}(1960)}]{Ziman-ElectronsPhonons-1960}%
  \BibitemOpen
  \bibfield  {author} {\bibinfo {author} {\bibfnamefont {J.~M.}\ \bibnamefont
  {Ziman}},\ }\href@noop {} {\emph {\bibinfo {title} {Electrons and Phonons:
  The Theory of Transport Phenomena in Solids}}}\ (\bibinfo  {publisher}
  {Oxford University Press},\ \bibinfo {year} {1960})\ pp.\ \bibinfo {pages}
  {264--298}\BibitemShut {NoStop}%
\bibitem [{\citenamefont
  {Srivastava}(1990)}]{Srivastava-PhysicsOfPhonons-1990}%
  \BibitemOpen
  \bibfield  {author} {\bibinfo {author} {\bibfnamefont {G.~P.}\ \bibnamefont
  {Srivastava}},\ }\href@noop {} {\emph {\bibinfo {title} {The Physics of
  Phonons}}}\ (\bibinfo  {publisher} {CRC Press},\ \bibinfo {year} {1990})\ p.\
  \bibinfo {pages} {122}\BibitemShut {NoStop}%
\bibitem [{\citenamefont {Dirac}(1958)}]{Dirac-PrinciplesOfQM-1958}%
  \BibitemOpen
  \bibfield  {author} {\bibinfo {author} {\bibfnamefont {P.}~\bibnamefont
  {Dirac}},\ }\href@noop {} {\emph {\bibinfo {title} {The Principles of Quantum
  Mechanics}}}\ (\bibinfo  {publisher} {Oxford University Press},\ \bibinfo
  {year} {1958})\ pp.\ \bibinfo {pages} {232--235}\BibitemShut {NoStop}%
\bibitem [{\citenamefont {Ward}\ \emph {et~al.}(2009)\citenamefont {Ward},
  \citenamefont {Broido}, \citenamefont {Stewart},\ and\ \citenamefont
  {Deinzer}}]{Ward-PhysRevB.80.125203-Tcond-2009}%
  \BibitemOpen
  \bibfield  {author} {\bibinfo {author} {\bibfnamefont {A.}~\bibnamefont
  {Ward}}, \bibinfo {author} {\bibfnamefont {D.~A.}\ \bibnamefont {Broido}},
  \bibinfo {author} {\bibfnamefont {D.~A.}\ \bibnamefont {Stewart}}, \ and\
  \bibinfo {author} {\bibfnamefont {G.}~\bibnamefont {Deinzer}},\ }\href
  {\doibase 10.1103/PhysRevB.80.125203} {\bibfield  {journal} {\bibinfo
  {journal} {Phys. Rev. B}\ }\textbf {\bibinfo {volume} {80}},\ \bibinfo
  {pages} {125203} (\bibinfo {year} {2009})}\BibitemShut {NoStop}%
\bibitem [{\citenamefont {Maradudin}\ and\ \citenamefont
  {Fein}(1962)}]{Maradudin-Fein-ScatteringOfNeutrons-1962}%
  \BibitemOpen
  \bibfield  {author} {\bibinfo {author} {\bibfnamefont {A.}~\bibnamefont
  {Maradudin}}\ and\ \bibinfo {author} {\bibfnamefont {A.}~\bibnamefont
  {Fein}},\ }\href@noop {} {\bibfield  {journal} {\bibinfo  {journal} {Phys.
  Rev.}\ }\textbf {\bibinfo {volume} {128}},\ \bibinfo {pages} {2589} (\bibinfo
  {year} {1962})}\BibitemShut {NoStop}%
\bibitem [{\citenamefont
  {Maradudin}(1962)}]{Maradudin-ThermalExpFreqShifts-1962}%
  \BibitemOpen
  \bibfield  {author} {\bibinfo {author} {\bibfnamefont {A.}~\bibnamefont
  {Maradudin}},\ }\href@noop {} {\bibfield  {journal} {\bibinfo  {journal}
  {phys. status solidi B}\ }\textbf {\bibinfo {volume} {2}},\ \bibinfo {pages}
  {1493} (\bibinfo {year} {1962})}\BibitemShut {NoStop}%
\bibitem [{\citenamefont {Paulatto}\ \emph {et~al.}(2015)\citenamefont
  {Paulatto}, \citenamefont {Errea}, \citenamefont {Calandra},\ and\
  \citenamefont {Mauri}}]{Paulatto-PhysRevB.91.054304_lifetimes-2015}%
  \BibitemOpen
  \bibfield  {author} {\bibinfo {author} {\bibfnamefont {L.}~\bibnamefont
  {Paulatto}}, \bibinfo {author} {\bibfnamefont {I.}~\bibnamefont {Errea}},
  \bibinfo {author} {\bibfnamefont {M.}~\bibnamefont {Calandra}}, \ and\
  \bibinfo {author} {\bibfnamefont {F.}~\bibnamefont {Mauri}},\ }\href
  {\doibase 10.1103/PhysRevB.91.054304} {\bibfield  {journal} {\bibinfo
  {journal} {Phys. Rev. B}\ }\textbf {\bibinfo {volume} {91}},\ \bibinfo
  {pages} {054304} (\bibinfo {year} {2015})}\BibitemShut {NoStop}%
\bibitem [{\citenamefont {Karttunen}\ \emph {et~al.}(2010)\citenamefont
  {Karttunen}, \citenamefont {Fässler}, \citenamefont {Linnolahti},\ and\
  \citenamefont {Pakkanen}}]{Karttunen-Structuralprinc-2010}%
  \BibitemOpen
  \bibfield  {author} {\bibinfo {author} {\bibfnamefont {A.~J.}\ \bibnamefont
  {Karttunen}}, \bibinfo {author} {\bibfnamefont {T.~F.}\ \bibnamefont
  {Fässler}}, \bibinfo {author} {\bibfnamefont {M.}~\bibnamefont
  {Linnolahti}}, \ and\ \bibinfo {author} {\bibfnamefont {T.~A.}\ \bibnamefont
  {Pakkanen}},\ }\href@noop {} {\bibfield  {journal} {\bibinfo  {journal}
  {Inorg. Chem.}\ }\textbf {\bibinfo {volume} {50}},\ \bibinfo {pages} {1733}
  (\bibinfo {year} {2010})}\BibitemShut {NoStop}%
\bibitem [{\citenamefont {Giannozzi}\ \emph {et~al.}(2009)\citenamefont
  {Giannozzi}, \citenamefont {Baroni}, \citenamefont {Bonini}, \citenamefont
  {Calandra}, \citenamefont {Car}, \citenamefont {Cavazzoni}, \citenamefont
  {Ceresoli}, \citenamefont {Chiarotti}, \citenamefont {Cococcioni},
  \citenamefont {Dabo}, \citenamefont {{Dal Corso}}, \citenamefont
  {de~Gironcoli}, \citenamefont {Fabris}, \citenamefont {Fratesi},
  \citenamefont {Gebauer}, \citenamefont {Gerstmann}, \citenamefont
  {Gougoussis}, \citenamefont {Kokalj}, \citenamefont {Lazzeri}, \citenamefont
  {Martin-Samos}, \citenamefont {Marzari}, \citenamefont {Mauri}, \citenamefont
  {Mazzarello}, \citenamefont {Paolini}, \citenamefont {Pasquarello},
  \citenamefont {Paulatto}, \citenamefont {Sbraccia}, \citenamefont {Scandolo},
  \citenamefont {Sclauzero}, \citenamefont {Seitsonen}, \citenamefont
  {Smogunov}, \citenamefont {Umari},\ and\ \citenamefont
  {Wentzcovitch}}]{QE-2009}%
  \BibitemOpen
  \bibfield  {author} {\bibinfo {author} {\bibfnamefont {P.}~\bibnamefont
  {Giannozzi}}, \bibinfo {author} {\bibfnamefont {S.}~\bibnamefont {Baroni}},
  \bibinfo {author} {\bibfnamefont {N.}~\bibnamefont {Bonini}}, \bibinfo
  {author} {\bibfnamefont {M.}~\bibnamefont {Calandra}}, \bibinfo {author}
  {\bibfnamefont {R.}~\bibnamefont {Car}}, \bibinfo {author} {\bibfnamefont
  {C.}~\bibnamefont {Cavazzoni}}, \bibinfo {author} {\bibfnamefont
  {D.}~\bibnamefont {Ceresoli}}, \bibinfo {author} {\bibfnamefont {G.~L.}\
  \bibnamefont {Chiarotti}}, \bibinfo {author} {\bibfnamefont {M.}~\bibnamefont
  {Cococcioni}}, \bibinfo {author} {\bibfnamefont {I.}~\bibnamefont {Dabo}},
  \bibinfo {author} {\bibfnamefont {A.}~\bibnamefont {{Dal Corso}}}, \bibinfo
  {author} {\bibfnamefont {S.}~\bibnamefont {de~Gironcoli}}, \bibinfo {author}
  {\bibfnamefont {S.}~\bibnamefont {Fabris}}, \bibinfo {author} {\bibfnamefont
  {G.}~\bibnamefont {Fratesi}}, \bibinfo {author} {\bibfnamefont
  {R.}~\bibnamefont {Gebauer}}, \bibinfo {author} {\bibfnamefont
  {U.}~\bibnamefont {Gerstmann}}, \bibinfo {author} {\bibfnamefont
  {C.}~\bibnamefont {Gougoussis}}, \bibinfo {author} {\bibfnamefont
  {A.}~\bibnamefont {Kokalj}}, \bibinfo {author} {\bibfnamefont
  {M.}~\bibnamefont {Lazzeri}}, \bibinfo {author} {\bibfnamefont
  {L.}~\bibnamefont {Martin-Samos}}, \bibinfo {author} {\bibfnamefont
  {N.}~\bibnamefont {Marzari}}, \bibinfo {author} {\bibfnamefont
  {F.}~\bibnamefont {Mauri}}, \bibinfo {author} {\bibfnamefont
  {R.}~\bibnamefont {Mazzarello}}, \bibinfo {author} {\bibfnamefont
  {S.}~\bibnamefont {Paolini}}, \bibinfo {author} {\bibfnamefont
  {A.}~\bibnamefont {Pasquarello}}, \bibinfo {author} {\bibfnamefont
  {L.}~\bibnamefont {Paulatto}}, \bibinfo {author} {\bibfnamefont
  {C.}~\bibnamefont {Sbraccia}}, \bibinfo {author} {\bibfnamefont
  {S.}~\bibnamefont {Scandolo}}, \bibinfo {author} {\bibfnamefont
  {G.}~\bibnamefont {Sclauzero}}, \bibinfo {author} {\bibfnamefont {A.~P.}\
  \bibnamefont {Seitsonen}}, \bibinfo {author} {\bibfnamefont {A.}~\bibnamefont
  {Smogunov}}, \bibinfo {author} {\bibfnamefont {P.}~\bibnamefont {Umari}}, \
  and\ \bibinfo {author} {\bibfnamefont {R.~M.}\ \bibnamefont {Wentzcovitch}},\
  }\href {http://www.quantum-espresso.org} {\bibfield  {journal} {\bibinfo
  {journal} {J. Phys.: Condens. Matter}\ }\textbf {\bibinfo {volume} {21}},\
  \bibinfo {pages} {395502 (19pp)} (\bibinfo {year} {2009})}\BibitemShut
  {NoStop}%
\bibitem [{\citenamefont {Garrity}\ \emph {et~al.}(2014)\citenamefont
  {Garrity}, \citenamefont {Bennett}, \citenamefont {Rabe},\ and\ \citenamefont
  {Vanderbilt}}]{Garrity-pseudopotentials-2014}%
  \BibitemOpen
  \bibfield  {author} {\bibinfo {author} {\bibfnamefont {K.~F.}\ \bibnamefont
  {Garrity}}, \bibinfo {author} {\bibfnamefont {J.~W.}\ \bibnamefont
  {Bennett}}, \bibinfo {author} {\bibfnamefont {K.~M.}\ \bibnamefont {Rabe}}, \
  and\ \bibinfo {author} {\bibfnamefont {D.}~\bibnamefont {Vanderbilt}},\
  }\href@noop {} {\bibfield  {journal} {\bibinfo  {journal} {Comput. Mater.
  Sci.}\ }\textbf {\bibinfo {volume} {81}},\ \bibinfo {pages} {446} (\bibinfo
  {year} {2014})}\BibitemShut {NoStop}%
\bibitem [{\citenamefont {Perdew}\ \emph {et~al.}(1996)\citenamefont {Perdew},
  \citenamefont {Burke},\ and\ \citenamefont
  {Ernzerhof}}]{Perdew-generalized-1996}%
  \BibitemOpen
  \bibfield  {author} {\bibinfo {author} {\bibfnamefont {J.~P.}\ \bibnamefont
  {Perdew}}, \bibinfo {author} {\bibfnamefont {K.}~\bibnamefont {Burke}}, \
  and\ \bibinfo {author} {\bibfnamefont {M.}~\bibnamefont {Ernzerhof}},\
  }\href@noop {} {\bibfield  {journal} {\bibinfo  {journal} {Phys. Rev. Lett.}\
  }\textbf {\bibinfo {volume} {77}},\ \bibinfo {pages} {3865} (\bibinfo {year}
  {1996})}\BibitemShut {NoStop}%
\bibitem [{\citenamefont {Baroni}\ \emph {et~al.}(2001)\citenamefont {Baroni},
  \citenamefont {de~Gironcoli}, \citenamefont {Dal~Corso},\ and\ \citenamefont
  {Giannozzi}}]{Baroni-RevModPhys.73.515-DFTP-2001}%
  \BibitemOpen
  \bibfield  {author} {\bibinfo {author} {\bibfnamefont {S.}~\bibnamefont
  {Baroni}}, \bibinfo {author} {\bibfnamefont {S.}~\bibnamefont
  {de~Gironcoli}}, \bibinfo {author} {\bibfnamefont {A.}~\bibnamefont
  {Dal~Corso}}, \ and\ \bibinfo {author} {\bibfnamefont {P.}~\bibnamefont
  {Giannozzi}},\ }\href {\doibase 10.1103/RevModPhys.73.515} {\bibfield
  {journal} {\bibinfo  {journal} {Rev. Mod. Phys.}\ }\textbf {\bibinfo {volume}
  {73}},\ \bibinfo {pages} {515} (\bibinfo {year} {2001})}\BibitemShut
  {NoStop}%
\bibitem [{\citenamefont {Li}\ \emph {et~al.}(2012{\natexlab{a}})\citenamefont
  {Li}, \citenamefont {Mingo}, \citenamefont {Lindsay}, \citenamefont {Broido},
  \citenamefont {Stewart},\ and\ \citenamefont
  {Katcho}}]{Li-Gaussian_PhysRevB.85.195436-2012}%
  \BibitemOpen
  \bibfield  {author} {\bibinfo {author} {\bibfnamefont {W.}~\bibnamefont
  {Li}}, \bibinfo {author} {\bibfnamefont {N.}~\bibnamefont {Mingo}}, \bibinfo
  {author} {\bibfnamefont {L.}~\bibnamefont {Lindsay}}, \bibinfo {author}
  {\bibfnamefont {D.~A.}\ \bibnamefont {Broido}}, \bibinfo {author}
  {\bibfnamefont {D.~A.}\ \bibnamefont {Stewart}}, \ and\ \bibinfo {author}
  {\bibfnamefont {N.~A.}\ \bibnamefont {Katcho}},\ }\href@noop {} {\bibfield
  {journal} {\bibinfo  {journal} {Phys. Rev. B}\ }\textbf {\bibinfo {volume}
  {85}},\ \bibinfo {pages} {195436} (\bibinfo {year}
  {2012}{\natexlab{a}})}\BibitemShut {NoStop}%
\bibitem [{\citenamefont {Li}\ \emph {et~al.}(2012{\natexlab{b}})\citenamefont
  {Li}, \citenamefont {Lindsay}, \citenamefont {Broido}, \citenamefont
  {Stewart},\ and\ \citenamefont
  {Mingo}}]{Li-ThirdOrderPy_PhysRevB.86.174307-2012}%
  \BibitemOpen
  \bibfield  {author} {\bibinfo {author} {\bibfnamefont {W.}~\bibnamefont
  {Li}}, \bibinfo {author} {\bibfnamefont {L.}~\bibnamefont {Lindsay}},
  \bibinfo {author} {\bibfnamefont {D.~A.}\ \bibnamefont {Broido}}, \bibinfo
  {author} {\bibfnamefont {D.~A.}\ \bibnamefont {Stewart}}, \ and\ \bibinfo
  {author} {\bibfnamefont {N.}~\bibnamefont {Mingo}},\ }\href@noop {}
  {\bibfield  {journal} {\bibinfo  {journal} {Phys. Rev. B}\ }\textbf {\bibinfo
  {volume} {86}},\ \bibinfo {pages} {174307} (\bibinfo {year}
  {2012}{\natexlab{b}})}\BibitemShut {NoStop}%
\bibitem [{\citenamefont {Nilsson}\ and\ \citenamefont
  {Nelin}(1972)}]{Nilsson-PhysRevB.6.3777_dispersion_exp_Si_alpha-1972}%
  \BibitemOpen
  \bibfield  {author} {\bibinfo {author} {\bibfnamefont {G.}~\bibnamefont
  {Nilsson}}\ and\ \bibinfo {author} {\bibfnamefont {G.}~\bibnamefont
  {Nelin}},\ }\href {\doibase 10.1103/PhysRevB.6.3777} {\bibfield  {journal}
  {\bibinfo  {journal} {Phys. Rev. B}\ }\textbf {\bibinfo {volume} {6}},\
  \bibinfo {pages} {3777} (\bibinfo {year} {1972})}\BibitemShut {NoStop}%
\bibitem [{\citenamefont {Yin}\ and\ \citenamefont
  {Cohen}(1982)}]{Yin-abInitioPhononsSi-1982}%
  \BibitemOpen
  \bibfield  {author} {\bibinfo {author} {\bibfnamefont {M.}~\bibnamefont
  {Yin}}\ and\ \bibinfo {author} {\bibfnamefont {M.~L.}\ \bibnamefont
  {Cohen}},\ }\href@noop {} {\bibfield  {journal} {\bibinfo  {journal} {Phys.
  Rev. B}\ }\textbf {\bibinfo {volume} {25}},\ \bibinfo {pages} {4317}
  (\bibinfo {year} {1982})}\BibitemShut {NoStop}%
\bibitem [{\citenamefont {Giannozzi}\ \emph {et~al.}(1991)\citenamefont
  {Giannozzi}, \citenamefont {de~Gironcoli}, \citenamefont {Pavone},\ and\
  \citenamefont {Baroni}}]{Giannozzi-abInitioPhononsSiGe-1991}%
  \BibitemOpen
  \bibfield  {author} {\bibinfo {author} {\bibfnamefont {P.}~\bibnamefont
  {Giannozzi}}, \bibinfo {author} {\bibfnamefont {S.}~\bibnamefont
  {de~Gironcoli}}, \bibinfo {author} {\bibfnamefont {P.}~\bibnamefont
  {Pavone}}, \ and\ \bibinfo {author} {\bibfnamefont {S.}~\bibnamefont
  {Baroni}},\ }\href@noop {} {\bibfield  {journal} {\bibinfo  {journal} {Phys.
  Rev. B}\ }\textbf {\bibinfo {volume} {43}},\ \bibinfo {pages} {7231}
  (\bibinfo {year} {1991})}\BibitemShut {NoStop}%
\bibitem [{\citenamefont {Wei}\ and\ \citenamefont
  {Chou}(1992)}]{Wei-abInitioPhononsSi-1992}%
  \BibitemOpen
  \bibfield  {author} {\bibinfo {author} {\bibfnamefont {S.}~\bibnamefont
  {Wei}}\ and\ \bibinfo {author} {\bibfnamefont {M.}~\bibnamefont {Chou}},\
  }\href@noop {} {\bibfield  {journal} {\bibinfo  {journal} {Phys. rev. lett.}\
  }\textbf {\bibinfo {volume} {69}},\ \bibinfo {pages} {2799} (\bibinfo {year}
  {1992})}\BibitemShut {NoStop}%
\bibitem [{\citenamefont {Favot}\ and\ \citenamefont
  {Dal~Corso}(1999)}]{Favot-abInitioPhononsSiGGA-1999}%
  \BibitemOpen
  \bibfield  {author} {\bibinfo {author} {\bibfnamefont {F.}~\bibnamefont
  {Favot}}\ and\ \bibinfo {author} {\bibfnamefont {A.}~\bibnamefont
  {Dal~Corso}},\ }\href@noop {} {\bibfield  {journal} {\bibinfo  {journal}
  {Phys. Rev. B}\ }\textbf {\bibinfo {volume} {60}},\ \bibinfo {pages} {11427}
  (\bibinfo {year} {1999})}\BibitemShut {NoStop}%
\bibitem [{\citenamefont {H{\"a}rk{\"o}nen}\ and\ \citenamefont
  {Karttunen}(2014)}]{Harkonen-NTE-2014}%
  \BibitemOpen
  \bibfield  {author} {\bibinfo {author} {\bibfnamefont {V.~J.}\ \bibnamefont
  {H{\"a}rk{\"o}nen}}\ and\ \bibinfo {author} {\bibfnamefont {A.~J.}\
  \bibnamefont {Karttunen}},\ }\href@noop {} {\bibfield  {journal} {\bibinfo
  {journal} {Phys. Rev. B}\ }\textbf {\bibinfo {volume} {89}},\ \bibinfo
  {pages} {024305} (\bibinfo {year} {2014})}\BibitemShut {NoStop}%
\bibitem [{\citenamefont
  {Inyushkin}(2002)}]{Inyushkin-dSiThermalConductivity-2002}%
  \BibitemOpen
  \bibfield  {author} {\bibinfo {author} {\bibfnamefont {A.}~\bibnamefont
  {Inyushkin}},\ }\href@noop {} {\bibfield  {journal} {\bibinfo  {journal}
  {Inorg. mater.}\ }\textbf {\bibinfo {volume} {38}},\ \bibinfo {pages} {427}
  (\bibinfo {year} {2002})}\BibitemShut {NoStop}%
\bibitem [{\citenamefont {Esfarjani}\ \emph {et~al.}(2011)\citenamefont
  {Esfarjani}, \citenamefont {Chen},\ and\ \citenamefont
  {Stokes}}]{Esfarjani-HeatTransportInSilicon-PhysRevB.84.085204-2011}%
  \BibitemOpen
  \bibfield  {author} {\bibinfo {author} {\bibfnamefont {K.}~\bibnamefont
  {Esfarjani}}, \bibinfo {author} {\bibfnamefont {G.}~\bibnamefont {Chen}}, \
  and\ \bibinfo {author} {\bibfnamefont {H.~T.}\ \bibnamefont {Stokes}},\
  }\href {\doibase 10.1103/PhysRevB.84.085204} {\bibfield  {journal} {\bibinfo
  {journal} {Phys. Rev. B}\ }\textbf {\bibinfo {volume} {84}},\ \bibinfo
  {pages} {085204} (\bibinfo {year} {2011})}\BibitemShut {NoStop}%
\bibitem [{\citenamefont {Inyushkin}\ \emph {et~al.}(2004)\citenamefont
  {Inyushkin}, \citenamefont {Taldenkov}, \citenamefont {Gibin}, \citenamefont
  {Gusev},\ and\ \citenamefont {Pohl}}]{Inyushkin-TCondSilicon-2004}%
  \BibitemOpen
  \bibfield  {author} {\bibinfo {author} {\bibfnamefont {A.~V.}\ \bibnamefont
  {Inyushkin}}, \bibinfo {author} {\bibfnamefont {A.~N.}\ \bibnamefont
  {Taldenkov}}, \bibinfo {author} {\bibfnamefont {A.~M.}\ \bibnamefont
  {Gibin}}, \bibinfo {author} {\bibfnamefont {A.~V.}\ \bibnamefont {Gusev}}, \
  and\ \bibinfo {author} {\bibfnamefont {H.-J.}\ \bibnamefont {Pohl}},\ }\href
  {\doibase 10.1002/pssc.200405341} {\bibfield  {journal} {\bibinfo  {journal}
  {Phys. Status Solidi (c)}\ }\textbf {\bibinfo {volume} {1}},\ \bibinfo
  {pages} {2995} (\bibinfo {year} {2004})}\BibitemShut {NoStop}%
\bibitem [{\citenamefont
  {Carruthers}(1961)}]{Carruthers-ThermalConductivityLowTemp-1961}%
  \BibitemOpen
  \bibfield  {author} {\bibinfo {author} {\bibfnamefont {P.}~\bibnamefont
  {Carruthers}},\ }\href@noop {} {\bibfield  {journal} {\bibinfo  {journal}
  {Rev. Mod. Phys.}\ }\textbf {\bibinfo {volume} {33}},\ \bibinfo {pages} {92}
  (\bibinfo {year} {1961})}\BibitemShut {NoStop}%
\bibitem [{\citenamefont {Shevelkov}\ \emph {et~al.}(2011)\citenamefont
  {Shevelkov}, \citenamefont {Kelm}, \citenamefont {Olenev}, \citenamefont
  {Kulbachinskii},\ and\ \citenamefont
  {Kytin}}]{Shevelkov-AnomalouslyLowTCond-2011}%
  \BibitemOpen
  \bibfield  {author} {\bibinfo {author} {\bibfnamefont {A.}~\bibnamefont
  {Shevelkov}}, \bibinfo {author} {\bibfnamefont {E.}~\bibnamefont {Kelm}},
  \bibinfo {author} {\bibfnamefont {A.}~\bibnamefont {Olenev}}, \bibinfo
  {author} {\bibfnamefont {V.}~\bibnamefont {Kulbachinskii}}, \ and\ \bibinfo
  {author} {\bibfnamefont {V.}~\bibnamefont {Kytin}},\ }\href@noop {}
  {\bibfield  {journal} {\bibinfo  {journal} {Semicond.}\ }\textbf {\bibinfo
  {volume} {45}},\ \bibinfo {pages} {1399} (\bibinfo {year}
  {2011})}\BibitemShut {NoStop}%
\bibitem [{\citenamefont {Lindsay}\ and\ \citenamefont
  {Broido}(2008)}]{Lindsay-ThreePhononPhaseSpace-2008}%
  \BibitemOpen
  \bibfield  {author} {\bibinfo {author} {\bibfnamefont {L.}~\bibnamefont
  {Lindsay}}\ and\ \bibinfo {author} {\bibfnamefont {D.}~\bibnamefont
  {Broido}},\ }\href@noop {} {\bibfield  {journal} {\bibinfo  {journal} {J.
  Phys.: Condens. Matter}\ }\textbf {\bibinfo {volume} {20}},\ \bibinfo {pages}
  {165209} (\bibinfo {year} {2008})}\BibitemShut {NoStop}%
\bibitem [{\citenamefont {Allen}\ and\ \citenamefont
  {Feldman}(1993)}]{Allen-ThermalCondGlass-1993}%
  \BibitemOpen
  \bibfield  {author} {\bibinfo {author} {\bibfnamefont {P.~B.}\ \bibnamefont
  {Allen}}\ and\ \bibinfo {author} {\bibfnamefont {J.~L.}\ \bibnamefont
  {Feldman}},\ }\href@noop {} {\bibfield  {journal} {\bibinfo  {journal} {Phys.
  Rev. B}\ }\textbf {\bibinfo {volume} {48}},\ \bibinfo {pages} {12581}
  (\bibinfo {year} {1993})}\BibitemShut {NoStop}%
\bibitem [{\citenamefont {Allen}\ \emph {et~al.}(1999)\citenamefont {Allen},
  \citenamefont {Feldman}, \citenamefont {Fabian},\ and\ \citenamefont
  {Wooten}}]{Allen-DiffusonsLoconsPropagons-1999}%
  \BibitemOpen
  \bibfield  {author} {\bibinfo {author} {\bibfnamefont {P.~B.}\ \bibnamefont
  {Allen}}, \bibinfo {author} {\bibfnamefont {J.~L.}\ \bibnamefont {Feldman}},
  \bibinfo {author} {\bibfnamefont {J.}~\bibnamefont {Fabian}}, \ and\ \bibinfo
  {author} {\bibfnamefont {F.}~\bibnamefont {Wooten}},\ }\href@noop {}
  {\bibfield  {journal} {\bibinfo  {journal} {Philos. Mag. B}\ }\textbf
  {\bibinfo {volume} {79}},\ \bibinfo {pages} {1715} (\bibinfo {year}
  {1999})}\BibitemShut {NoStop}%
\end{thebibliography}%
\end{document}